\documentclass[fleqn,usenatbib]{mnras}

\usepackage{newtxtext,newtxmath}
\usepackage[T1]{fontenc}

\DeclareRobustCommand{\VAN}[3]{#2}
\let\VANthebibliography\thebibliography
\def\thebibliography{\DeclareRobustCommand{\VAN}[3]{##3}\VANthebibliography}

\usepackage{graphicx}	% Including figure files
\usepackage{amsmath}	% Advanced maths commands
\usepackage{float}
\usepackage{orcidlink}
\usepackage{longtable} % For long tables
\usepackage{pdflscape}   % For rotating specific pages in PDF output
\usepackage{rotating}
\usepackage{booktabs}
\usepackage{siunitx}

\renewcommand{\%}{\text{ per cent}}

\newcommand{\oii}{[O\,\textsc{II}]}
\newcommand{\oiii}{[O\,\textsc{III}]}
\newcommand{\nii}{[N\,\textsc{II}]}

\newcommand{\feiii}{[Fe\,\textsc{III}]}

\newcommand{\cliii}{[Cl\,\textsc{III}]}
\newcommand{\siii}{[S\,\textsc{III}]}
\newcommand{\sii}{[S\,\textsc{II}]}
\newcommand{\ariv}{[Ar\,\textsc{IV}]}
\newcommand{\hi}{H\,\textsc{I}}
\newcommand{\hii}{H\,\textsc{II}}

\newcommand{\oiirls}{O\,\textsc{II}}

\title[The DESIRED temperature-metallicity relations]{The DESIRED temperature–metallicity relations in star-forming regions: probing the Galactic radial and azimuthal metallicity distributions}

\author[I. Rafael Martínez-Hernández et al.]{
I. Rafael Martínez-Hernández$^{1}$\thanks{E-mail: isaias.martinez@unah.hn}\orcidlink{0009-0003-8206-6678}, 
J. Eduardo Méndez-Delgado$^{2}$ \thanks{E-mail: jmendez@astro.unam.mx}\orcidlink{0000-0002-6972-6411}, 
César Esteban$^{3,4}$\orcidlink{0000-0002-5247-5943},
\and
Jorge García-Rojas$^{3,4}$\orcidlink{0000-0002-6138-1869},
Leticia Carigi$^{2}$\orcidlink{0000-0002-2023-466X}, 
Luis F. Rodríguez$^{5}$\orcidlink{0000-0003-2737-5681},
Luis A. Zapata$^{5}$\orcidlink{0000-0003-2343-7937},
\and
F. Fabián Rosales-Ortega$^{6}$\orcidlink{0000-0002-3642-9146},
Maialen Orte-García$^{3,4}$\orcidlink{0000-0002-0539-1720},
Elena Reyes-Rodríguez$^{3,4}$\orcidlink{0000-0003-1192-6987},
Karla Z. Arellano-Córdova$^{7}$\orcidlink{0000-0002-2644-3518},
\and
Kathryn Kreckel$^{8}$\orcidlink{0000-0001-6551-3091},
Natascha Sattler$^{8}$\orcidlink{0000-0002-8883-6018},
Christophe Morisset$^{9}$\orcidlink{0000-0001-5801-6724}, 
Manuel Peimbert$^{2}$,
Silvia Torres-Peimbert$^{2}$\orcidlink{0000-0001-6775-8615},
\and
Miriam Peña$^{2}$\orcidlink{0009-0007-5891-420X},
\v{Z}ofia Chrob\'akov\'a$^{10}$\orcidlink{0000-0002-9895-6638}, 
Eleonora Zari$^{11}$\orcidlink{0000-0003-3769-8812}, 
David A. Espinoza-Galeas$^{1}$\orcidlink{0000-0003-2971-0439}
\\
$^{1}$Departamento de Astronom\'ia y Astrof\'isica, Facultad de Ciencias Espaciales, Universidad Nacional Aut\'onoma de Honduras, Bulevar Suyapa, Tegucigalpa,\\ M.D.C, Honduras, Centroamérica\\
$^{2}$Instituto de Astronom\'{\i}a, Universidad Nacional Aut\'onoma de M\'exico, A.P. 70-264, 04510 CDMX, M\'exico\\
$^{3}$Instituto de Astrof\'isica de Canarias, E-38205, La Laguna, Tenerife, Spain\\
$^{4}$Departamento de Astrof\'isica, Universidad de La Laguna, E-38206 La Laguna, Tenerife, Spain\\
$^{5}$Instituto de Radioastronomía y Astrofísica, Universidad Nacional Autónoma de México, C.P. 58089 Morelia, Michoacán, México\\
$^{6}$Instituto Nacional de Astrofísica, Óptica y Electrónica (INAOE SECIHTI), Luis E. Erro 1, 72840, Tonantzintla, Puebla, México\\
$^{7}$Institute for Astronomy, University of Edinburgh, Royal Observatory, Edinburgh, EH9 3HJ, United Kingdom\\
$^{8}$Astronomisches Rechen-Institut, Zentrum f\"{u}r Astronomie der Universit\"{a}t Heidelberg, M\"{o}nchhofstra\ss e 12-14, D-69120 Heidelberg, Germany\\
$^{9}$ Universidad Nacional Aut\'onoma de M\'exico, Instituto de Ciencias F\'sicas, Av. Universidad s/n, 62210 Cuernavaca, Mor., M\'exico\\
$^{10}$Institute of Construction and Architecture, Slovak Academy of Science, 84503 Bratislava, Slovakia\\
$^{11}$Dipartimento di Fisica e Astronomia, Universit{\`a} degli Studi di Firenze, Via G. Sansone 1, I-50019, Sesto F.no (Firenze), Italy
}

\date{Accepted XXX. Received YYY; in original form ZZZ}

\pubyear{\the\year{}}

\begin{document}
\label{firstpage}
\pagerange{\pageref{firstpage}--\pageref{lastpage}}
\maketitle


\begin{abstract}
We analyse a sample of 225 star-forming regions from the DESIRED-E project, each with simultaneous determinations of the electron temperature from ionized nitrogen and oxygen, $T_{\rm e}$(\nii) and $T_{\rm e}$(\oiii), respectively. We derive new empirical relations connecting the gas-phase metallicity to the global electron temperature, $T_{\rm e}$(H$^+$), as determined via radio observations. We establish two calibrations: one assuming a homogeneous temperature distribution ($t^2 = 0$, the ``direct method''), and another accounting for internal temperature fluctuations ($t^2 > 0$). Applying these calibrations to 460 radio observations of Galactic \hii~regions spanning Galactocentric distances from $\sim0.1$ to 16 kpc, we determine the radial O/H gradient in the Milky Way under both assumptions. We further compare these nebular gradients to independent metallicity estimates from young O- and B-type stars and Cepheid variables. We find that the $t^2 > 0$ calibration yields a gradient in excellent agreement with stellar-based determinations, whereas the $t^2 = 0$ method underestimates metallicities by up to $\sim$0.3 dex. This discrepancy cannot be reconciled by invoking oxygen depletion onto dust grains or nucleosynthetic processing via the CNO cycle in massive stars. We also find that one widely used relation in the literature, assuming $t^2 = 0$, produces an excessively steep gradient --likely due to the use of outdated atomic data and pre-CCD observations. Finally, we explore potential azimuthal variations in the Galactic metallicity distribution driven by the presence of the spiral arms, finding no evidence for variations larger than $\sim$0.1 dex with respect to the general radial gradient.% This result highlights the high degree of mixing of the ISM across a large portion of the Galactic disc.
%We study the behavior of O abundances in the Milky Way in a sample of more than 200 deep optical spectra of HII regions and star-forming galaxies (SFGs) of the local Universe exploring the impact of (1): the 

%This is a simple template for authors to write new MNRAS papers.  
%The abstract should briefly describe the aims, methods, and main results of the paper.
%It should be a single paragraph not more than 250 words (200 words for Letters).
%No references should appear in the abstract.
\end{abstract}


%%%%%%%%%%%%%%%%%%%%%%%%%%%%%%%%%%%%%%%%%%%%%%%%%%

%%%%%%%%%%%%%%%%% BODY OF PAPER %%%%%%%%%%%%%%%%%%

\section{Introduction}
\label{Sec:Intro}

In star-forming ionised nebulae, metallicity and electron temperature are closely linked. Heavy elements such as oxygen play a key role in the thermal balance of the gas. On the one hand, owing to their atomic structure --prone to collisional excitations-- they govern a significant fraction of the gas cooling \citep{Osterbrock:06}.
On the other hand, the ionising stars, which dominate the heating processes, tend to be cooler as the metallicity increases \citep{Vilchez:88a}. 

At radio wavelengths, the nebular continuum in \hii~regions is dominated by thermal free-free radiation (bremsstrahlung), and its ratio to the brightness temperature of a hydrogen recombination line provides a powerful diagnostic of the gas electron temperature, $T_{\rm e}$ \citep{Churchwell:75}. As a result, it is feasible to estimate the metallicity distribution in galaxies such as the Milky Way from radio observations of star-forming regions \citep{Shaver:83, Quireza:06, Wenger:19}, overcoming the limitations imposed by Galactic extinction on optical studies \citep{peimbert:78,Deharveng:00,ArellanoCordova:20,ArellanoCordova:21,MendezDelgado:22}, and even those affecting infrared observations \citep{Rudolph:06, Pineda:24}.

To fully exploit the potential of radio-wavelength observations of \hii~regions for studying the global gas metallicity, it is necessary to establish an accurate calibration between metallicity -- traced by the O/H ratio of ionised nebulae -- and $T_{\rm e}$. This approach assumes that metal-line cooling is primarily driven by oxygen, which accounts for approximately 55\% of the total metal content \citep{Peimbert:07}, and that the cooling contributions from other heavy elements are correlated with that of oxygen. Accurate determination of the oxygen-based metallicity requires precise measurements of the emission fluxes from its various ionic transitions, together with a reliable global electron density and temperature determinations. The former can be achieved through observations of bright collisionally excited lines (CELs) or ultra-faint heavy-element recombination lines (RLs). The latter can be obtained using multiple line diagnostics or through the nebular continuum.

However, for decades, a discrepancy between nebular metallicities inferred from CELs and RLs has been known, with the former being systematically lower \citep{Bowen:39,Wyse:42}. This inconsistency, known as the abundance discrepancy problem and quantified by the Abundance Discrepancy Factor (ADF) -- defined as the logarithmic difference between both estimates \citep{Liu:00}-- has been attributed to various physical phenomena, including the presence of $T_{\rm e}$ inhomogeneities in the nebular gas \citep{Peimbert:67, Peimbert:69}, which could also explain the discrepancies observed among some $T_{\rm e}$ diagnostics \citep{Esteban:04}. Since the seminal work of \citet{Peimbert:67}, there has been an ongoing debate regarding the presence or absence of such internal variations \citep{Perez:97, GarciaRojas:07b, Stasinska:13, Chen:23, RickardsVaught:24, MendezDelgado:23a, MendezDelgado:24, MendezDelgado:25}. If $T_{\rm e}$ fluctuations are present and not properly accounted for, such inhomogeneities can systematically bias most heavy-element abundance determinations in ionised interstellar gas based on optical spectra using CELs, which remain the most accessible observational tracers of metallicity both in the local Universe and at high redshift \citep{ArellanoCordova:22, MendezDelgado:23preprint,Cameron:23,Rogers:24,Marconi:24,Cataldi:25}.

The abundance discrepancy problem implies the existence of at least two distinct calibrations of the $T_{\rm e}$–metallicity relation for photoionised star-forming regions. One assumes thermally homogeneous conditions, while the other considers $T_{\rm e}$ fluctuations using the mathematical formalism of \citet{Peimbert:67} and the empirical relations reported by \citet{MendezDelgado:23a}. In this work, we derive these relations and apply them to a large sample of 460 Galactic \hii~regions observed at radio wavelengths, with $T_{\rm e}$ determinations, to obtain the metallicity gradient of the Galaxy under both assumptions on the $T_{\rm e}$ structure. We then compare these distributions with independent estimates of the current metallicity of the Galactic local interstellar medium (ISM), such as chemical abundances measured in the photospheres of young O- and B-type stars or in classical Cepheids whose chemical composition has not been significantly altered. This approach allows us to test the plausibility of $T_{\rm e}$  inhomogeneities in star-forming regions and to establish the $T_{\rm e}$-metallicity relation that provides the best agreement across different tracers of the present day metallicity of the Galactic ISM.

This work is organised as follows: Section~\ref{Sec:Observational_Sample} describes the observational datasets used in this study, including optical spectra of \hii~regions, radio measurements, and stellar abundance determinations. Section~\ref{sec:Methods} details the methodology adopted to derive homogeneous Galactocentric distances, electron temperatures, and oxygen abundances under different assumptions about the thermal structure of the nebulae. In Section~\ref{Sec:Results}, we present the new empirical temperature–metallicity relations obtained from the combined optical and radio samples. In Section~\ref{Sec:gradients}, we apply these calibrations to derive the Galactic radial metallicity gradient, while in Section~\ref{sec:nebular_vs_stellar} we compare the nebular results with those from young stars and Cepheids. In Section~\ref{Sec:Azimuthal}, we investigate the possible existence of azimuthal metallicity variations driven by spiral arms across the Galactic disk. Finally, Section~\ref{Sec:conclusions} summarises our main results and conclusions.

\section{Observational data }
\label{Sec:Observational_Sample}

This work makes use of optical and radio observations of both Galactic and extragalactic star-forming regions collected from the literature. Additionally, oxygen abundance (O/H) determinations from O- and B-type stars and Cepheids available in the literature are used for comparison with the nebular results. 

\begin{figure}
    \includegraphics[width=1.\columnwidth]{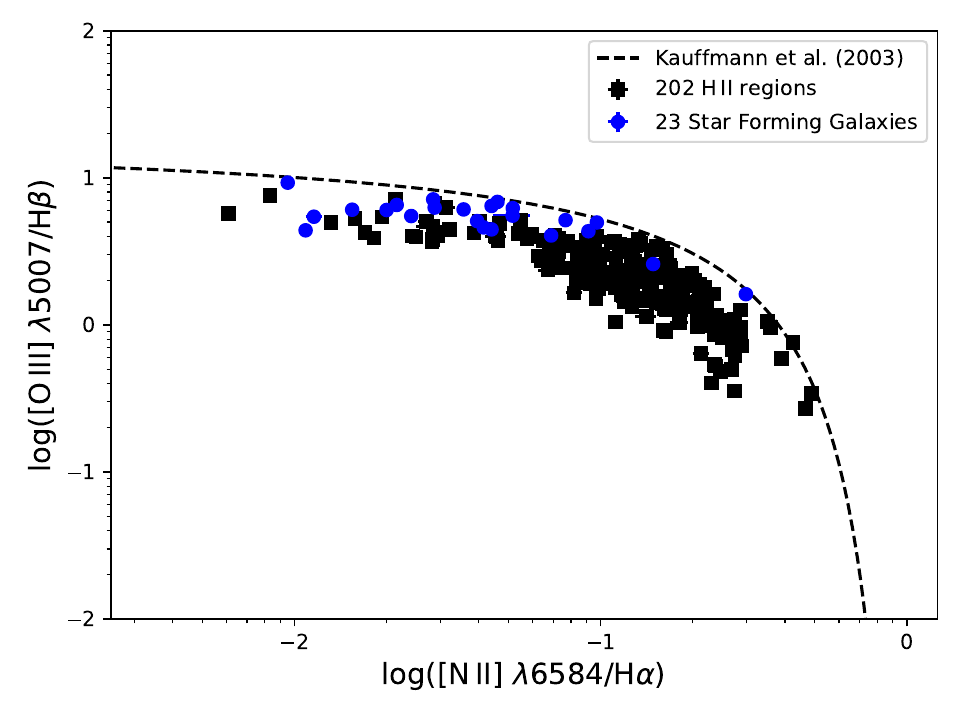}
    \caption{BPT diagram \citep{Baldwin:81} of the selected nebular optical spectra. The dashed curve represents the empirical demarcation by \citet{Kauffmann:03}. Objects located above the curve correspond to regions dominated by harder ionizing radiation fields, while those below it are associated with softer, star-formation–driven ionization and other soft ionizing sources.}
    \label{fig:BPT}
\end{figure}

For the optical observations of star-forming regions, we use the sample presented in Table~1 of the supplementary data, which is part of the DEep Spectra of Ionized REgions Database Extended (DESIRED-E) project \citep{MendezDelgado:23b, MendezDelgado:24b, Esteban:25}. This project compiles published deep optical spectra of ionised nebulae, enabling the detection of faint emission lines, particularly auroral lines sensitive to $T_{\rm e}$, such as \oiii~$\lambda 4363$, \siii~$\lambda 6312$, and/or \nii~$\lambda 5755$, with uncertainties below 40\%. The subsample selected for this work consists of 225 spectra of Galactic and extragalactic star-forming regions in which both \nii~$\lambda 5755$ and \oiii~$\lambda 4363$ lines are detected. Our selected sample is limited to the local Universe, with redshifts $z < 0.1$. Their location in the Baldwin–Phillips–Terlevich (BPT) diagram \citep{Baldwin:81} is shown in Fig.~\ref{fig:BPT}. This diagram allows the separation of regions according to the hardness of their ionising radiation and is commonly used to distinguish between regions ionised by massive O- or early B-type stars and those dominated by harder or alternative ionisation mechanisms. These include, but are not limited to, Active Galactic Nuclei (AGN), hot evolved stars (e.g. white dwarfs or post-AGB stars), shocks associated with outflows or jets, as well as diffuse ionised gas, whose dominant ionisation sources are still under debate  \citep{Kumari:19}. In this work, we distinguish between star-forming galaxies and individual \hii~regions by considering that, in the former, the \hii~regions are not spatially resolved.

The selection of these 225 optical spectra allows us to consider the complete $T_{\rm e}$ structure of ionised nebulae across a wide range of metallicities, from $12+\text{log(O/H)}\sim7.5$ to $12+\text{log(O/H)}\sim 9.0$, encompassing both the highly ionised inner zones and the low-ionisation outer zones. This study enables a direct comparison between the $T_{\rm e}$ values derived from optical observations, which trace different ionisation layers of the gas, and those obtained from radio observations, which primarily trace the ionised hydrogen that occupies the entire volume of the nebula that emits the spectrum (see Section~\ref{Subsec:det_tem_metal}). The treatment of gas at intermediate ionisation degrees is implicitly accounted for by the adopted low- and high-ionisation temperature diagnostics. A detailed and explicit analysis of intermediate-ionisation temperature structures is presented by \citet{Orte:25}, to which we refer the reader for a comprehensive discussion and physical justification.

Ideally, one would adopt only optical observations of Galactic \hii~regions that fully cover the parameter space shown in Fig.~\ref{fig:BPT} and include simultaneous detections of both \nii~$\lambda5755$ and \oiii~$\lambda4363$. However, this is not feasible with current observations owing to the high optical extinction in the inner Galaxy -- where the highest-metallicity regions are located \citep{ArellanoCordova:21} -- and the low optical surface brightness of the most distant regions in the Galactic anticentre, which cover the lowest-metallicity regime \citep{Esteban:17}. Dedicated mapping projects of the Milky Way using optical integral field spectroscopy (IFS), such as the SDSS-V Local Volume Mapper (LVM) \citep{Drory:24, Kreckel:24, Kollmeier:25}, will enable broader coverage of the parameter space across the Galaxy, but they will likely still fall short of spanning the full range accessible through radio observations.

On the other hand, we adopt the $T_{\rm e}$ values of Galactic star-forming regions reported by radio studies based on 8–10 GHz observations from \citet{Quireza:06a} and \citet{Wenger:19}, obtained with the National Radio Astronomy Observatory (NRAO) 140 Foot (43 m) radio telescope and the Very Large Array (VLA), respectively. We also include data from \citet{Khan:24}, obtained in the 4–8 GHz band with the VLA. The 460 selected regions have kinematic distances derived using different rotation curves or parallax-based determinations from bright, compact objects such as methanol and water masers \citep{Khan:24}. These distances have been homogenised as described in detail in Section~\ref{Subsec:revis_distances} to allow a direct comparison between different data sets. \citet{MendezDelgado:22} demonstrated that there is good agreement between optical distance determinations based on Gaia parallaxes of ionising stars \citep{GaiaEDR3, Bailer-Jones:21} and kinematic distances derived using the rotation curve of \citet{Reid:14b,Reid:19}, which permits a meaningful comparison between optical and radio data in terms of Galactocentric position.

Finally, we also adopt the O/H abundance ratios derived for B-type stars by \citet{Nieva:12} and \citet{Wessmayer:2022}, as well as for O-type stars from \citet{Aschenbrenner:23}. In addition, we include the oxygen abundances reported by \citet{Luck:18} for Cepheid stars. It is possible to compare the resulting oxygen abundance gradients from \hii~regions with those obtained from these types of stars, since they have very short lifespans and their oxygen content is not expected to be significantly altered by nucleosynthetic processes \citep{Luck:08}. These stars have Gaia Early Data Release 3 (EDR3) parallaxes, and their distances were obtained from the photogeometric estimates of \citet{Bailer-Jones:21}.

\section{Methods}
\label{sec:Methods}

\subsection{Determining ${T_{\mathrm{e}}}$(H$^{+}$)-metallicity relations}
\label{Subsec:det_tem_metal}

Ionised nebulae exhibit a natural stratification in the ionisation state of the gas. In \hii~regions, the zones closer to the ionising sources are typically dominated by higher ionisation states, populated by ions such as O$^{2+}$, while lower ionisation states such as O$^+$ or N$^+$ prevail in the outer zones. For the vast majority of these objects, the contribution from higher ionisation states is negligible, even in the most metal-poor environments where the mean ionisation degree tends to be higher \citep{Izotov:99a, Berg:21, DominguezGuzman:22}. For this reason, diagnostics based on line intensity ratios of heavy elements are representative of specific volumes within the ionised gas. In contrast, $T_{\rm e}$ values derived from the ratio of the nebular continuum to a hydrogen recombination line, commonly measured at radio wavelengths, are representative of the entire ionised gas, since hydrogen is present throughout all ionised volumes, being denoted as ${T_{\mathrm{e}}}$(H$^{+}$). As shown by \citet{Peimbert:00}, it is possible to connect these $T_{\rm e}$ diagnostics using equation~\eqref{eq:TH}:

\begin{equation}
    \label{eq:TH}
    T_{\rm e}(\text{H}^{+})\approx\frac{T_{\rm e}(\text{O}^{+})\times n(\text{O}^{+}) +T_{\rm e}(\text{O}^{2+}) \times n(\text{O}^{2+})}{n(\text{O}^{+})+n(\text{O}^{2+})} \text{ [K],}
\end{equation}

\noindent where $T_{\rm e}(\text{X}^{+i})$ represents the $T_{\rm e}$ of the volume where the $\text{X}^{+i}$ ion coexists, and $n(\text{X}^{+i})$ denotes the abundance of that ion, X$^{+i}$/H$^+$.

Now, while equation~\eqref{eq:TH} is theoretically straightforward, its practical application encounters several challenges. To first order, $T_{\rm e}(\text{O}^{+}) \approx T_{\rm e}(\text{\oii~}\lambda\lambda7330^+/\lambda\lambda3728^+)$, and $T_{\rm e}(\text{N}^{+}) \approx T_{\rm e}(\text{\nii~}\lambda5755/\lambda6584)$, and photoionisation models predict that $T_{\rm e}(\text{O}^{+}) \approx T_{\rm e}(\text{N}^{+})$. However, empirical determinations typically show that $T_{\rm e}(\text{\oii~}\lambda\lambda7330^+/\lambda\lambda3728^+) \geq T_{\rm e}(\text{\nii~}\lambda5755/\lambda6584)$ \citep{Rogers:21, Zurita:21, MendezDelgado:23b,RickardsVaught:24}. These discrepancies are significant, on the order of several thousand degrees, which has important implications for the derived ionic abundances, particularly for $\text{O}^{+}/\text{H}^{+}$. \citet{MendezDelgado:23b} concluded that $T_{\rm e}(\text{\nii~}\lambda5755/\lambda6584)$ is a more robust $T_{\rm e}$ diagnostic for the low-ionisation gas volume, while $T_{\rm e}(\text{\oii~}\lambda\lambda7330^+/\lambda\lambda3728^+)$ may be biased by internal variations in the electron density, $n_{\rm e}$ \citep[see also][] {Kreckel:25}. The presence of such $n_{\rm e}$ variations has been widely demonstrated since several pioneering studies \citep{Seaton:57, Peimbert:71, Rubin:89}, although their effect remains an active topic of academic debate \citep{Chen:23, MendezDelgado:24}.

On the other hand, it is possible to assume $T_{\rm e}(\text{O}^{2+}) \approx T_{\rm e}(\text{\oiii~}\lambda4363/\lambda5007)$, which forms the fundamental basis of the so-called ``direct method'' \citep{Dinerstein:90, Peimbert:17}. However, in the presence of $T_{\rm e}$ inhomogeneities, this observational diagnostic can be biased towards higher $T_{\rm e}$  values than the actual average, owing to the exponential dependence of the emissivities of CELs on $T_{\rm e}$ \citep{Osterbrock:06}. In fact, several studies have shown that $T_{\rm e}(\text{\oiii~}\lambda4363/\lambda5007) \geq T_{\rm e}(\text{\oiirls~}\lambda \lambda 4649^+/\text{\oiii~}\lambda 5007)$ \citep{Peimbert:03, Esteban:04, GarciaRojas:07b, MendezDelgado:23a}, where the latter combines \oiirls~RLs with \oiii~CELs, even though both diagnostics probe the same ion, O$^{2+}$. In principle, under a homogeneous temperature structure, both diagnostics should yield equivalent results. In practice, this discrepancy is a manifestation of the ADF, as discussed in Section~\ref{Sec:Intro}. Fortunately, \citet{MendezDelgado:23a} established an empirical relation between $T_{\rm e}(\text{\nii~}\lambda5755/\lambda6584)$ and $T_{0}(\text{O}^{2+})$, the average $T_{\rm e}$ of the gas following the mathematical formalism of \citet{Peimbert:67}, which reconciles the abundances derived from \oiii~CELs and \oiirls~RLs.

Therefore, it is possible to establish a relation between $T_{\rm e}$ and the oxygen abundance under two assumptions: i) the presence of a homogeneous thermal structure in the ionised gas, as commonly assumed in the so-called ``direct method''; or ii) a relation that accounts for $T_{\rm e}$ inhomogeneities. In the latter case, we use the mathematical formalism of \citet{Peimbert:67} along with the empirical relations from \citet{MendezDelgado:23a}. $T_{\rm e}$ inhomogeneities can be modelled using the $t^2$ parameter, which represents the root mean square deviation from the average nebular $T_{\rm e}$ and provides a quantitative measure of the internal $T_{\rm e}$ variations within the gas. The abundances determined with $t^2 > 0$ are consistent with those derived from optical O recombination lines (O-RLs) \citep{Peimbert:03, Esteban:04, GarciaRojas:07b}.

In the case of a homogeneous thermal structure, i.e. $t^2 = 0$, equation~\eqref{eq:TH} becomes:

\begin{equation}
\label{eq:TH_t20}
T_{\rm e}(\text{H}^{+})_{t^2=0} \approx \frac{T_{\rm e}(\text{\nii}) \times n(\text{O}^{+}) + T_{\rm e}(\text{\oiii}) \times n(\text{O}^{2+})}{n(\text{O}^{+}) + n(\text{O}^{2+})} \text{ [K],}
\end{equation}

Whereas for the case where temperature variations are present, i.e. $t^2 > 0$, the expression becomes:

\begin{equation}
\label{eq:TH_t2geq0}
T_{\rm e}(\text{H}^{+})_{t^2>0} \approx \frac{T_{\rm e}(\text{\nii}) \times n(\text{O}^{+}) + T_{0}(\text{O}^{2+}) \times n(\text{O}^{2+})}{n(\text{O}^{+}) + n(\text{O}^{2+})} \text{ [K],}
\end{equation}

where $T_{0}(\text{O}^{2+})$ can be estimated using equation~(4) from \citet{MendezDelgado:23a}, presented here in Eq.~\eqref{eq:t22023}:

\begin{equation}
\label{eq:t22023}
T_{0}(\text{O}^{2+}) = \left( 1.17\pm 0.05 \right) \times T_{\rm e}(\text{\nii}) - \left(3340 \pm 470\right) \text{ [K].}
\end{equation}

Now, the choice of different $T_{\rm e}$ diagnostics has a significant impact on the ionic abundances derived from optical CELs. In contrast, the adopted value of $n_{\rm e}$ has a much smaller effect when using these same type of lines \citep{MendezDelgado:23b}. Note that $T_{\rm e}$(H$^+$), determined either with equation \eqref{eq:TH_t20} or \eqref{eq:TH_t2geq0}, does not necessarily differ significantly, since it is weighted by ionic abundances that vary consistently with the adopted electron temperature (a higher $T_{\rm e} (\text{O}^{2+})$ implies a lower $n(\text{O}^{2+})$ abundance). However, the ionic abundances do change substantially, thereby altering the temperature–metallicity relation.

The determination of physical conditions, specifically $n_{\rm e}$ and $ T_{\rm e}$, as well as the ionic abundances, was carried out using \textit{PyNeb} version 1.1.18 \citep{Luridiana:15, Morisset:20}, with the atomic data specified in Table~\ref{table:atomic_data} and the \hi~effective recombination coefficients from \citet{Storey:95}. The analysis followed the same methodology described in the DESIRED series of papers \citep{MendezDelgado:23b,MendezDelgado:24b,MendezDelgado:25,ReyesRodriguez:24,Orte:25, Esteban:25}.

In brief, $n_{\rm e}$ was determined using the available $n_{\rm e}$-sensitive diagnostics, such as \oii~$\lambda3727/\lambda3729$, \sii~$\lambda6731/\lambda6716$, \cliii~$\lambda5538/\lambda5518$, \feiii~$\lambda4658/\lambda4702$, and \ariv~$\lambda4740/\lambda4711$, through Monte Carlo simulations with 100 realizations to propagate the uncertainties. A representative value of $n_{\rm e}$ was then adopted following the criteria outlined in Section~5 of \citet{MendezDelgado:23b}. If $n_{\rm e}$(\sii~$\lambda6731/\lambda6716) < 100\ \text{cm}^{-3}$, we adopt $n_{\rm e} = 100^{+100} _{-99}\ \text{cm}^{-3}$.
If $100\ \text{cm}^{-3} \leq n_{\rm e}$(\sii~$\lambda6731/\lambda6716) < 1000\ \text{cm}^{-3}$, we adopt the average between $n_{\rm e}$(\sii~$\lambda6731/\lambda6716$) and $n_{\rm e}$(\oii~$\lambda3726/\lambda3729$). If $n_{\rm e}$(\sii~$\lambda6731/\lambda6716) \geq 1000\ \text{cm}^{-3}$, we adopt the average of the densities derived from \sii, \oii, \cliii, \feiii, and \ariv\ diagnostics. In cases where no density diagnostic is available, we adopt $n_{\rm e} = 100^{+100} _{-99}\ \text{cm}^{-3}$.

With the adopted average $n_{\rm e}$, we determined $T_{\rm e}$(\nii~$\lambda5755/\lambda6584$) and $T_{\rm e}$(\oiii~$\lambda4363/\lambda5007$). We then estimated the O$^+$/H$^+$ ratio using \oii~$\lambda\lambda3727+3729$, assuming the adopted $n_{\rm e}$ and $T_{\rm e}$(\nii~$\lambda5755/\lambda6584$). When \oii~$\lambda\lambda3727+3729$ lines were not available due to the spectral coverage, the \oii~$\lambda\lambda7320+7330$ lines were used instead. The difference in the ionic abundances between both line sets is up to $\sim$0.1 dex \citep{MendezDelgado:23b}, which is acceptable for the purposes of this work. The O$^{2+}$/H$^+$ ratio was determined using \oiii~$\lambda\lambda4959+5007$. In the case of $t^2=0$, we adopted $T_{\rm e}$(\oiii~$\lambda4363/\lambda5007$) to infer this abundance, while for $t^2>0$, we used $T_0$(O$^{2+}$). The total oxygen abundance, O/H, was then obtained directly as the sum of the ionic abundances: (O$^{+}$ + O$^{2+}$)/H$^+$. Details of the quality control applied to the DESIRED-E sample can be found in \citet{MendezDelgado:24b}.

\subsection{Analysis of the distance to Galactic \hii~regions}
\label{Subsec:revis_distances}

Distances to Galactic \hii~regions can be determined using several methods. The main ones include direct parallax-based distance determinations of their stellar populations  \citep{Hirota:07, Zhang:09, Reid:14a, MendezDelgado:20, MendezDelgado:22, Shen:25}, spectrophotometric methods \citep{Crampton:75, Russeil:03, Moises:11}, and kinematic estimates \citep{Quireza:06, Wenger:18}. Among these, the most accurate are the direct parallax-based determinations, in which stars or compact objects associated with the nebulae are observed with high-precision astrometry. However, most distant \hii~regions lack parallax measurements, and only kinematic distances are available.

Kinematic distances are determined using the radial velocity relative to the local standard of rest ($V_{\text{LSR}}$) of the nebular regions, in combination with a Galactic rotation model (GRM). Most rotation curves typically assume circular motion around the Galactic centre \citep{Clemens:85,Brand:93}. Naturally, every GRM yields two possible solutions for a given $V_{\text{LSR}}$: the ``near'' and ``far'' distances, a degeneracy known as the kinematic distance ambiguity (KDA). Additional information is therefore required to assign a unique distance to the source. Several methods have been developed to resolve the KDA for \hii~regions, including the Bayesian maximum likelihood method \citep{Reid:16, Reid:19}, H$_2$CO absorption \citep{Watson:03,Sewilo:04}, \hi~emission/absorption \citep{Kolpack:03,Anderson:09,Anderson:12,Khan:24}, and the \hi~self-absorption method \citep{Roman:09,Khan:24}.

Most distance determinations for the radio-observed \hii~regions in the adopted sample are based on kinematic methods. However, these determinations are not homogeneous in terms of the adopted GRM and other key parameters, and they may be affected by systematic biases. In contrast, the selected stellar objects and most Galactic \hii~regions with optical observations have distances derived from Gaia parallaxes. \citet{MendezDelgado:22} found that distances based on Gaia parallaxes and those obtained using the Monte Carlo kinematic method of \citet{Wenger:18}, adopting the GRM proposed by \citet{Reid:14b}, show good consistency. Therefore, a re-analysis of the kinematic distances is required to ensure consistency between the different datasets.

\citet{Quireza:06} derived heliocentric and Galactocentric distances ($d$ and $R_G$, respectively) for 111 \hii~regions using the observed $V_{\text{LSR}}$. For sources located within the solar orbit, they adopted the Galactic rotation curve of \citet{Clemens:85}, while for sources outside the solar circle, the rotation model of \citet{Brand:93} was used. These determinations assumed the standard Galactic parameters of $R_0 = 8.5$ kpc for the Galactocentric distance of the Sun and $\Theta_0 = 220$ km s$^{-1}$ for the circular velocity at the solar radius. Importantly, the calculations did not account for uncertainties in either the measured velocities or the model parameters. 

In the study by \citet{Wenger:19}, a total of 167 Galactic \hii~regions with reliable $T_{\rm e}$(H$^{+}$) measurements were considered. Of these, 46 have distances adopted from maser parallaxes reported by \citet{Reid:14b}, while the remaining 121 have kinematic distances derived from their radial velocities. To compute the kinematic distances, the authors used the GRM from \citet{Reid:14b}, adopting $R_0 \approx 8.34$ kpc and $\Theta_0 \approx 240$ km s$^{-1}$. This method explicitly incorporates uncertainties in both the measured velocities and the adopted Galactic parameters, ensuring that the resulting kinematic distance estimates properly reflect these uncertainties.

\citet{Khan:24} derived $d$ and $R_G$ for 220 \hii~regions using a combination of parallax-based measurements, literature kinematic distances with resolved ambiguities, and new kinematic determinations. For 28 sources, distances were adopted from maser parallax measurements by \citet{Reid:19}. An additional 121 regions were assigned literature-based kinematic distances in which the KDA had already been resolved by \citet{Anderson:09}, \citet{Anderson:12} and \citet{Urquhart:18}, and 13 objects in the Cygnus X complex were assumed to lie at a fixed distance of 1.4 kpc. For the remaining 53 regions, $V_{\text{LSR}}$ values were used together with the GRM of \citet{Reid:14b}, assuming $R_0 = 8.34$ kpc and $\Theta_0 = 240$ km s$^{-1}$. Kinematic distances derived in this way carry typical uncertainties of approximately 17\%, primarily due to streaming motions and model-dependent effects.

To reanalyse the kinematic distances of the sample in a homogeneous manner, we adopted the Monte Carlo technique described by \citet{Wenger:18}, using their publicly available \textit{KDUtils} software package \citep{Wenger:17} hosted on \textit{GitHub} \footnote{https://github.com/tvwenger/kd}. We adopted a solar Galactocentric distance of $R_0 = 8.2 \pm 0.1$ kpc, which is the most precisely determined value according to \citet{Bland:16} and is consistent with the location of the Galactic centre black hole \citep{GRAVITY:19}. In this model, we adopted the GRM of \citet{Reid:14b}, properly propagating the uncertainties from the $V_{\text{LSR}}$ measurements reported by the original authors. We also adopted a local circular rotation speed of $\Theta_0 \approx 240$ km s$^{-1}$ \citep{Reid:14b}. We used equations (4), (5), and (6) from \citet{Wenger:18} to correct the LSR velocities for the updated solar motion parameters. Each reanalysed region yields two possible distances due to the KDA. To resolve this ambiguity, we selected the solution closest to the reference value reported in the literature.

\section{Results}
\label{Sec:Results}

\subsection{The DESIRED $T_{\rm e}$–metallicity relations}
\label{Subsec:resulting_temperature_relations}

\begin{figure}
    \includegraphics[width=1.\columnwidth]{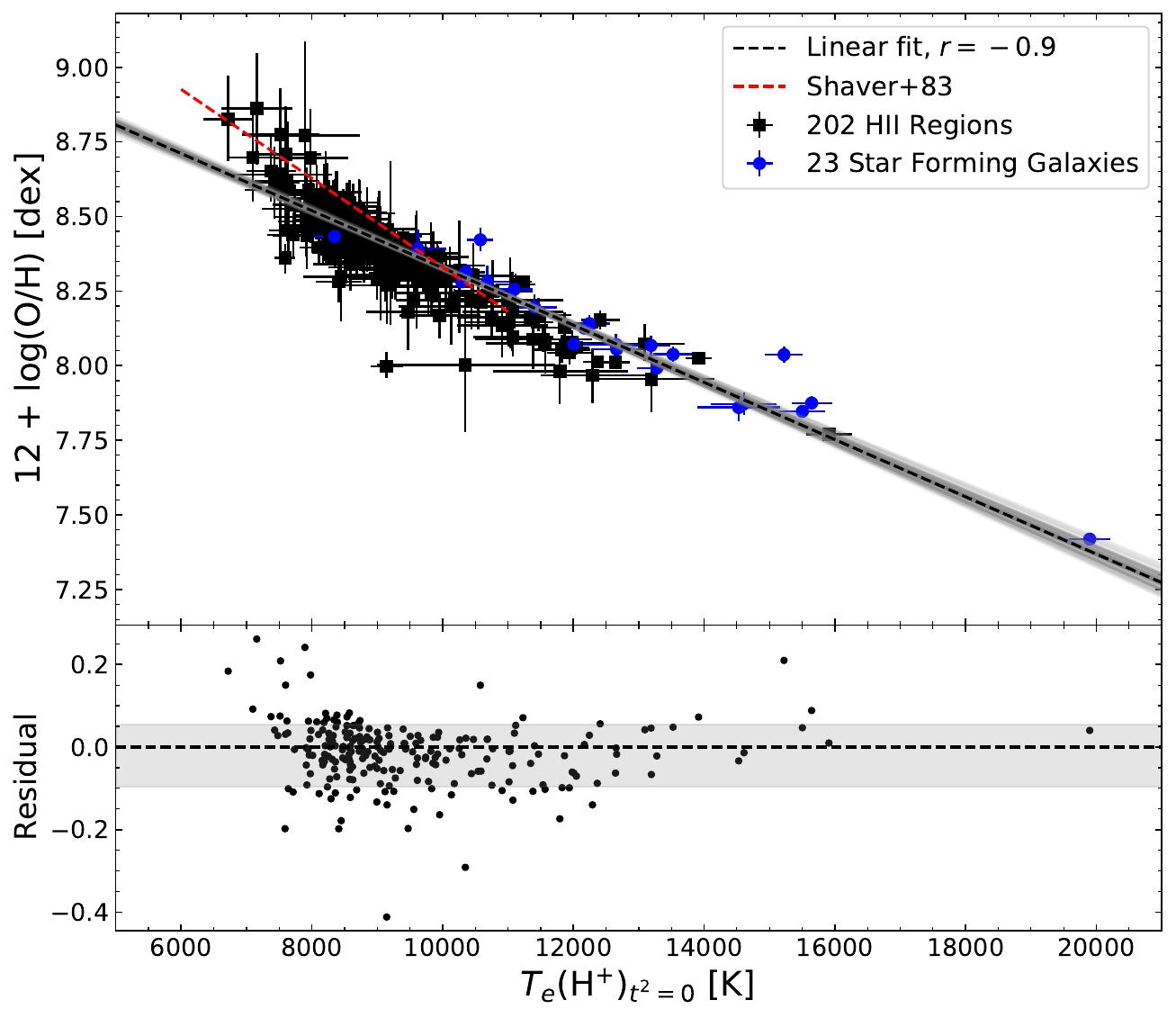}
    \caption{Relation between the total O/H abundance ratio, a proxy for the gas-phase metallicity, and the average gas $T_{\rm e}$, equivalent to the $T_{\rm e}$ of ionised hydrogen in the nebula ($T_{\rm e}$(H$^+$)) for our sample objects. The determination is based on equation~\eqref{eq:TH_t20}, which assumes a homogeneous $T_{\rm e}$ structure ($t^2 = 0$), and on the optical observational sample shown in Fig.~\ref{fig:BPT}. The upper panel shows the positions of the different observed objects along with the corresponding linear fit. $r$ is the Pearson correlation coefficient of the linear fit. As a comparison, the relation from \citet{Shaver:83}, which is typically used in the literature for radio observations, is shown in red. This latter relation has been plotted only over the temperature range used in the calibration of the original paper ($\sim$6000–11000 K). The lower panel displays the residuals relative to our best linear fit. The shaded gray region indicates the typical fitting uncertainty $1\sigma$, corresponding to +0.05 and $-$0.1.}
    \label{fig:metal_tem_t2eq0}
\end{figure}

\begin{figure}
    \includegraphics[width=1.\columnwidth]{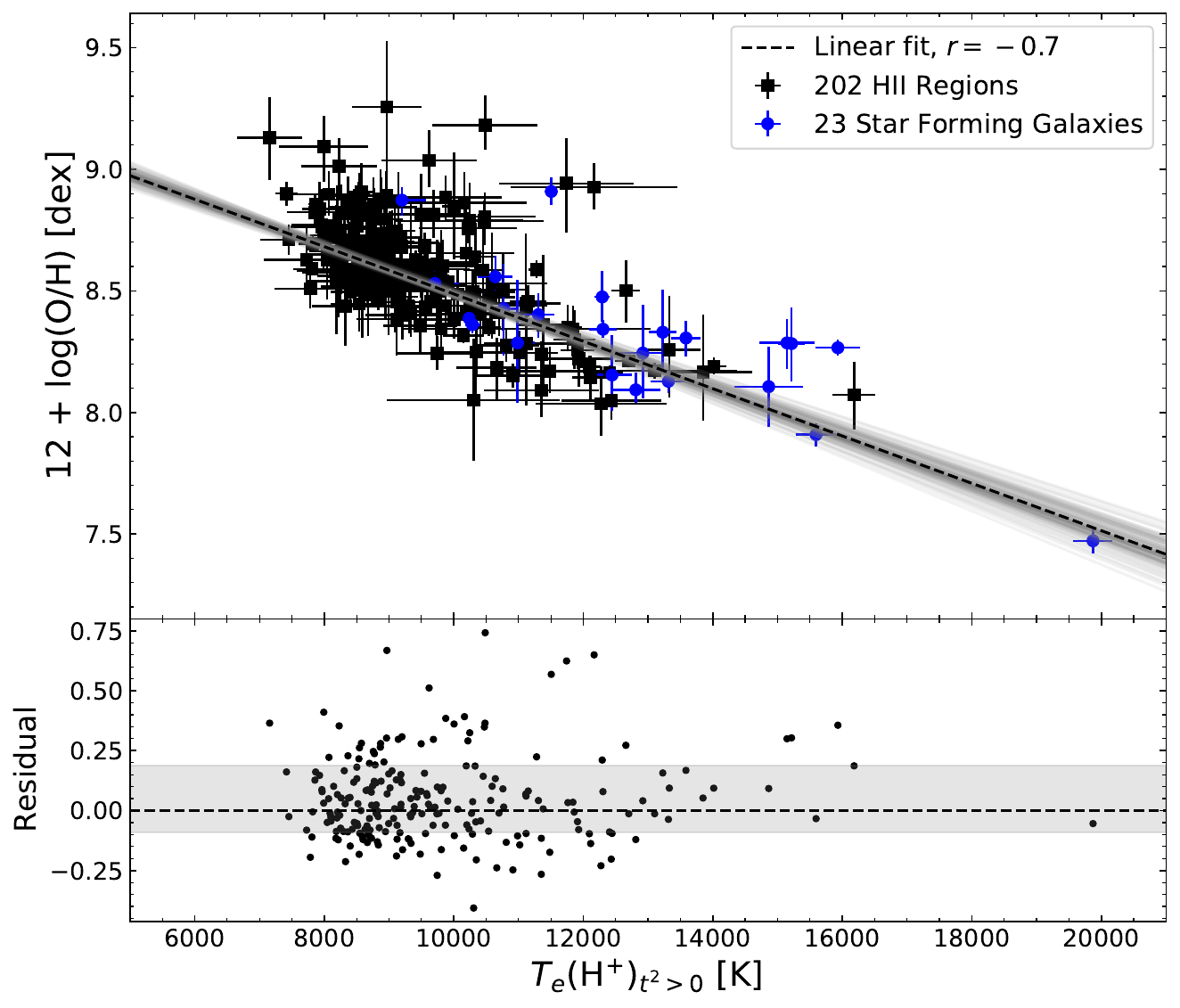}
    \caption{Same relation as in Fig.~\ref{fig:metal_tem_t2eq0}, but assuming an inhomogeneous temperature structure ($t^2>0$) as described in equation~\eqref{eq:TH_t2geq0}. The shaded gray region indicates the typical fitting uncertainty, corresponding to +0.2 and $-$0.1.}
    \label{fig:metal_tem_t2geq0}
\end{figure}

Following the methodology presented in Section~\ref{Subsec:det_tem_metal}, we determined the relation between $T_{\rm e}$ and the global metallicity of the ionised gas under two different assumptions: one assuming a homogeneous $T_{\rm e}$ structure and another that accounts for internal $T_{\rm e}$ variations. In Fig.~\ref{fig:metal_tem_t2eq0}, we show the relation between the oxygen abundance and the average $T_{\rm e}$ for the case of $t^2 = 0$, while Fig.~\ref{fig:metal_tem_t2geq0} presents the analogous relation for the $t^2 > 0$ case. In both cases, the fit was performed using the orthogonal distance regression (ODR) method, which accounts for uncertainties in both axes. Interestingly, the ODR fits are nearly identical to those obtained using a simple linear regression. It should be remembered that Fig.~\ref{fig:metal_tem_t2eq0} includes all our sample objects, Galactic and extragalactic \hii~regions as well as star-forming galaxies.

For the case of $t^2 = 0$, the resulting fit is:

\begin{equation}
\label{eq:tem_metal_t2eq0}
12 + \log(\text{O}/\text{H})_{t^2=0} = (9.29 \pm 0.02) - (0.96 \pm 0.02) \times \left( \dfrac{T_{\rm e}(\text{H}^{+})}{10^4 \text{K}}\right) \text{ [dex]},
\end{equation}

\noindent while for the case of $t^2 > 0$, the fit is:

\begin{equation}
\label{eq:tem_metal_t2gt0}
12 + \log(\text{O}/\text{H})_{t^2>0} = (9.46 \pm 0.05) - (0.97 \pm 0.05) \times \left( \dfrac{T_{\rm e}(\text{H}^{+})}{10^4 \text{K}}\right) \text{ [dex]}.
\end{equation}

Equations~\eqref{eq:tem_metal_t2eq0} and~\eqref{eq:tem_metal_t2gt0} are valid over a $T_{\rm e}$(H$^{+}$) range between approximately 6000 K to 20000 K, covering a metallicity range from 12 + log(O/H) $\sim 7.5$ to $\sim 9$. However, the empirical constraints at the high-temperature end are limited. In particular, only one object in our sample lies beyond $T_{\rm e} \sim 16,000$~K, at $T_{\rm e} \sim 20,000$~K. Therefore, the applicability of these relations at the highest temperatures should be interpreted with caution. These empirical relations are the most robust currently available in the literature, surpassing previous analyses such as that of \citet{Shaver:83} in both the number and quality of the spectra. It is important to highlight that, for each of these regions, both temperature diagnostics are available: those associated with high- and low-ionisation ions. This is extremely difficult to achieve observationally, as detecting \nii~$\lambda 5755$ -- a fundamental component of our analysis and directly dependent on the $T_{\rm e}$ of the low-ionisation gas -- is particularly challenging in metal-poor star-forming galaxies (SFGs) \citep[e.g.,][]{MendezDelgado:23a}. 

Interestingly, the 23 SFGs included in our sample follow the same $T_{\rm e}$–metallicity relation defined by Galactic and extragalactic \hii~regions. A separate linear fit using only the SFGs yields a slope of $\left(0.89 \pm 0.04 \right)\times10^{-4}$ for the $t^2 = 0$ case, and a slope of $\left(1.03 \pm 0.16 \right)\times10^{-4}$ for the $t^2 >0$ case, in good agreement with the slope derived from the full sample. This consistency suggests that the derived empirical relation may be applicable in extragalactic environments. The root-mean-square (rms) scatter of the residuals around the best-fit relation is 0.06 dex for the $t^2 = 0$ case, and 0.08 dex for the $t^2 > 0$ case. These values were computed as the standard deviation of the orthogonal distances from each data point to the best-fitting line. The larger scatter in the $t^2 > 0$ relation is expected, as it reflects both the observational uncertainties and the increased intrinsic dispersion introduced by the $t^2$ parameter.

We note that a small subset of seven regions at the high-metallicity end of the sample ($T_{\rm e} \lesssim 8000$~K) lies systematically above the linear temperature–metallicity relation shown in Figures~\ref{fig:metal_tem_t2eq0} and~\ref{fig:metal_tem_t2geq0} and appears to be more consistent with the calibration of \citet{Shaver:83}. We explored the use of a second-degree polynomial to test whether it was able to account for this behaviour; however, the difference between the linear and quadratic fits over the temperature range most densely populated by the data ($7000 \lesssim T_{\mathrm{e}} \lesssim 16,000$~K) is very small, reaching at most $\sim$0.05~dex. At the low-metallicity end, the quadratic fit departs significantly from the observational constraint at $T_{\rm e} \sim 20,000$~K, with deviations of up to $\sim$0.15~dex, potentially introducing systematic biases in the inferred abundances. Given the limited number of objects at high temperatures (or lower metallicities), we therefore adopt a linear relation as a more robust and conservative representation of the present data. This result underscores the need for deeper observations at low metallicities, with simultaneous detections of $T_{\rm e}$(\nii) and $T_{\rm e}$(\oiii), to better constrain the functional form of the temperature–metallicity relation.

\subsection{Galactocentric distances of Galactic \hii~regions}
\label{Subsec:Galactocentric}

\begin{figure}
    \includegraphics[width=1.\columnwidth]{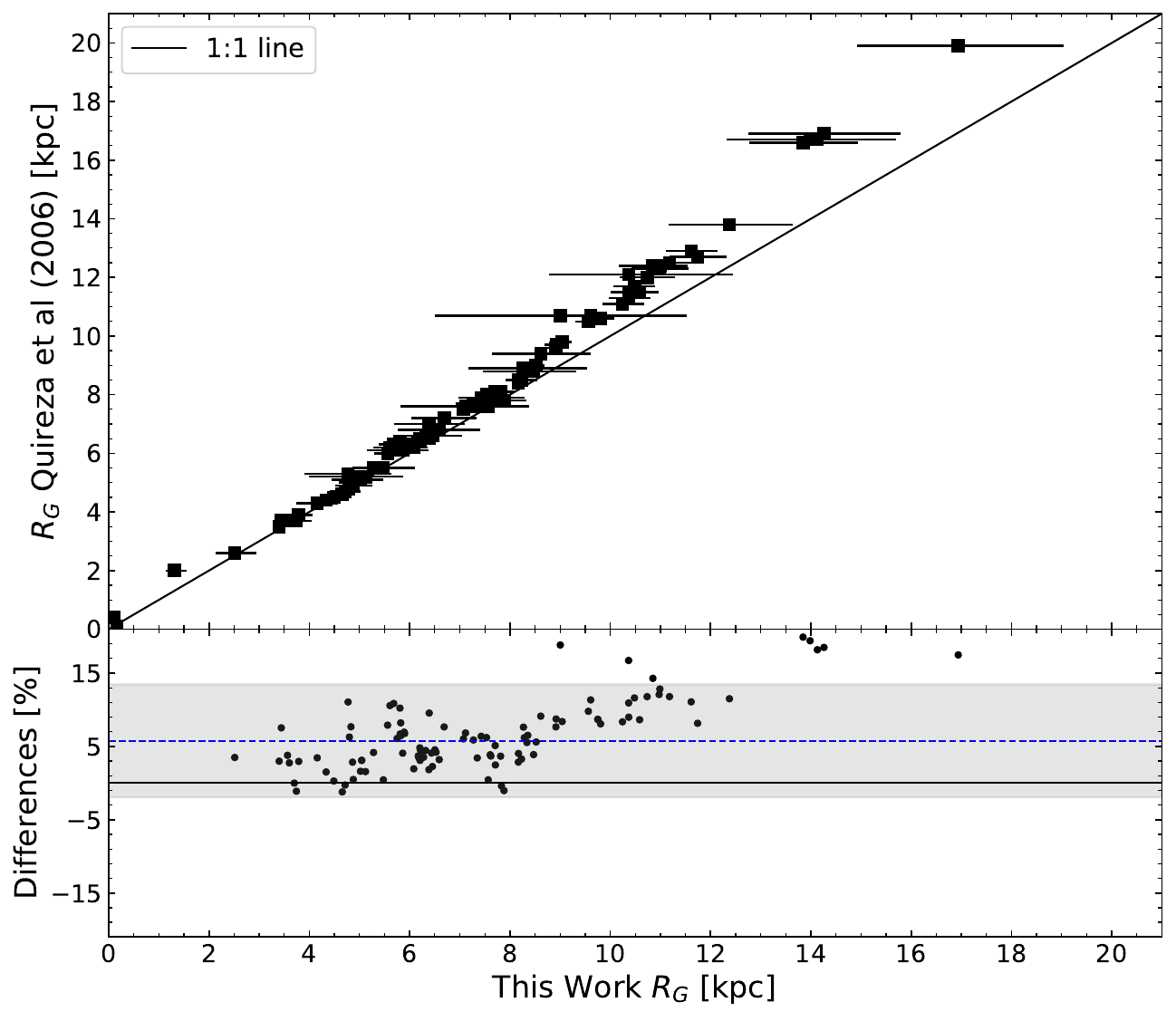}
    \caption{Comparison between the Galactocentric distances determined by \citet{Quireza:06} and those obtained in this work for Galactic \hii~regions (see Section~\ref{Subsec:revis_distances}). The upper panel shows the object-by-object comparison, while the lower panel displays the differences in percentage. The shaded gray region indicates the typical uncertainty $1\sigma$ confidence interval, corresponding to $+13$\% and $-2$\% and the blue dashed line represents the median with a value of +6\%.}
    \label{fig:Comparison_Quireza_Rgal}
\end{figure}

\begin{figure}
    \includegraphics[width=1.\columnwidth]{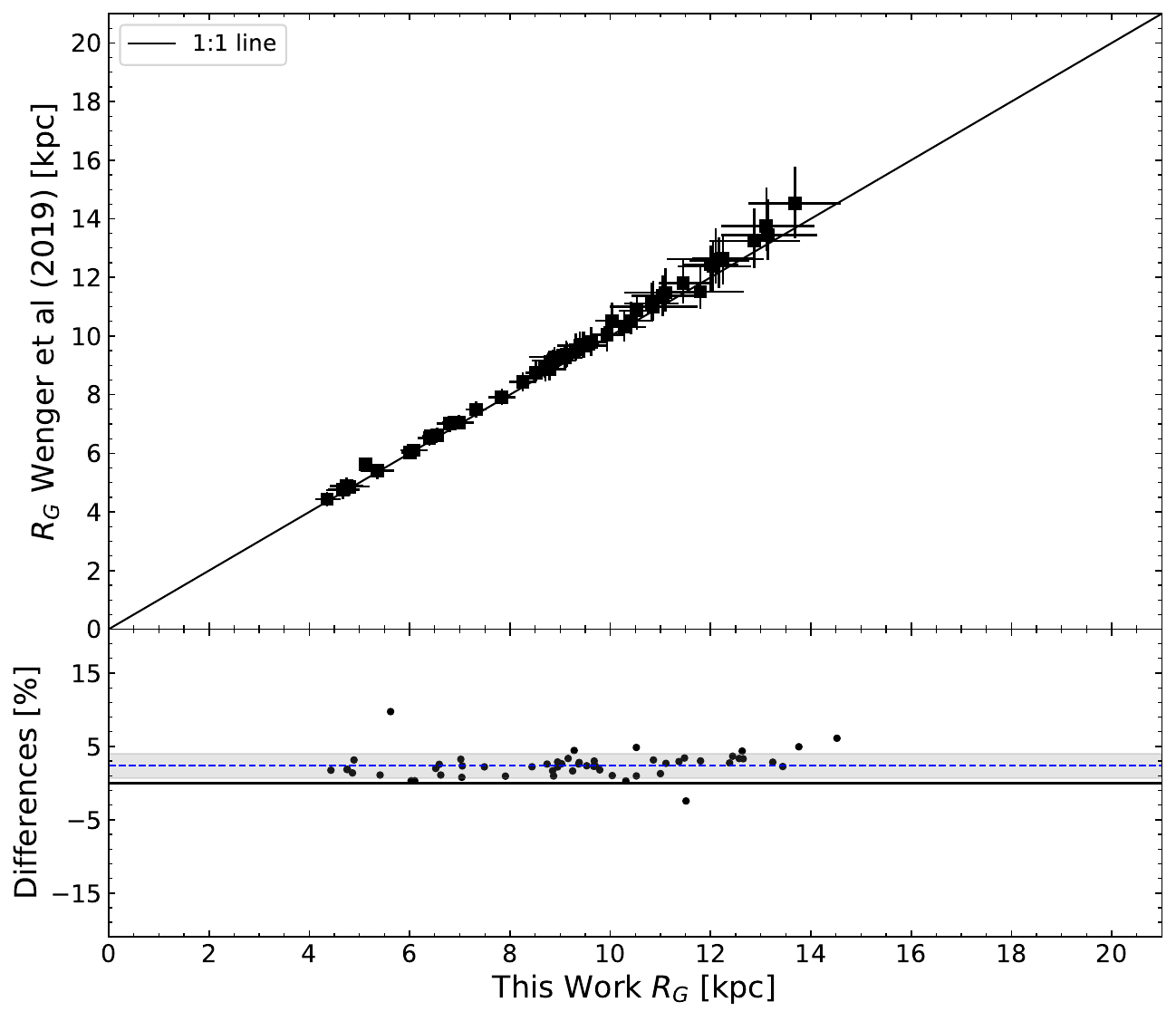}
    \caption{Analogous comparison to Fig.~\ref{fig:Comparison_Quireza_Rgal}, but using the data from \citet{Wenger:19}. The shaded gray region indicates the typical uncertainty $1\sigma$ confidence interval, corresponding to $+4$\% and $+1$\% and the blue dashed line represents the median with a value of $+2$\%.}
    \label{fig:Comparison_Wenger_Rgal}
\end{figure}

\begin{figure}
    \includegraphics[width=1.\columnwidth]{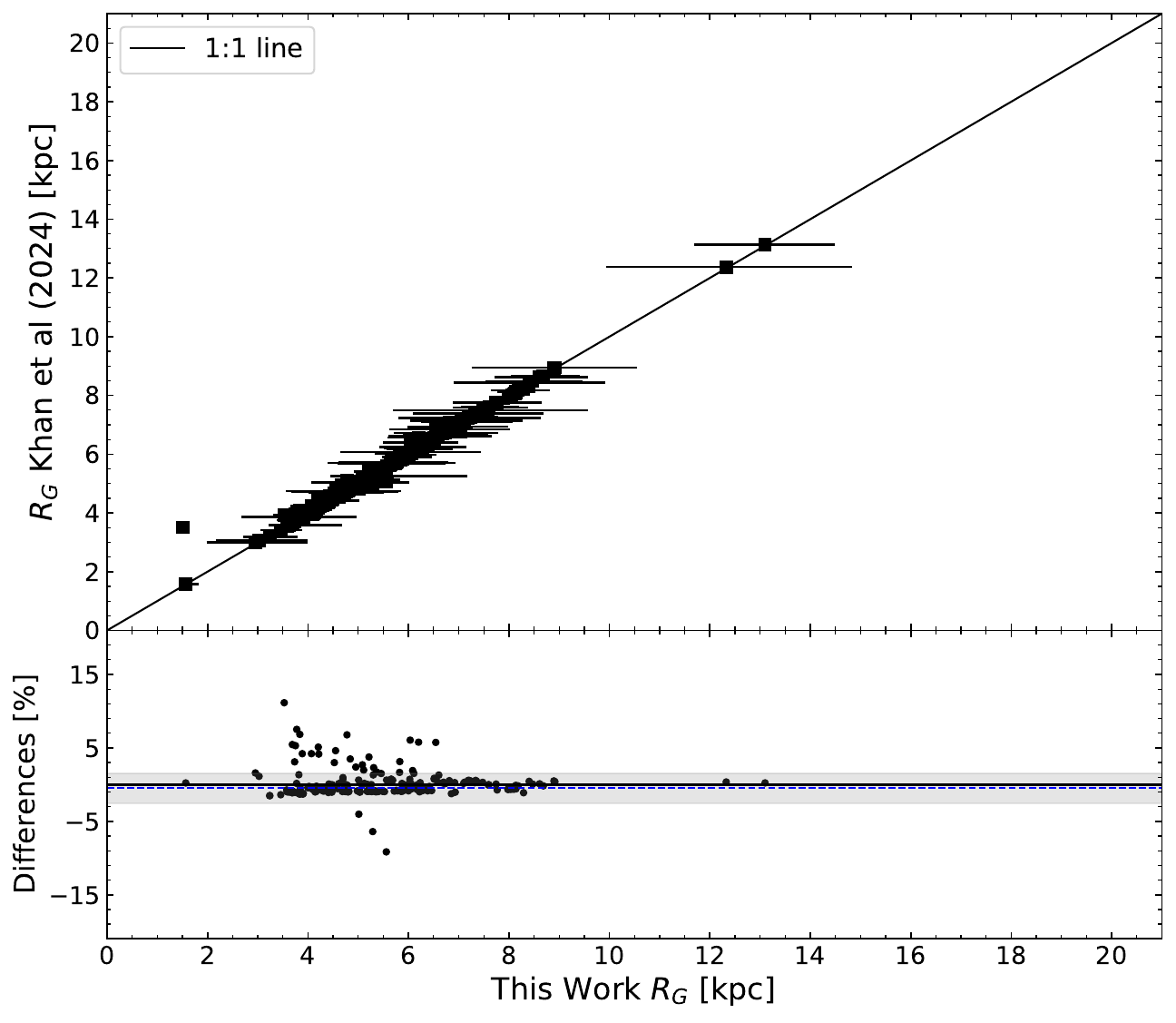}
    \caption{Analogous comparison to Fig.~\ref{fig:Comparison_Quireza_Rgal}, but using the data from \citet{Khan:24}. The shaded gray region indicates the typical uncertainty $1\sigma$ confidence interval, corresponding to $+2$\% and $-2$\% and the blue dashed line represents the median with a value of $-1$\%.}
    \label{fig:Comparison_Khan_Rgal}
\end{figure}

Following the procedure described in Section~\ref{Subsec:revis_distances}, we determined $d$ and $R_G$ from the kinematic distance estimates in a homogeneous manner. In Figs.~\ref{fig:Comparison_Quireza_Rgal}, \ref{fig:Comparison_Wenger_Rgal}, and \ref{fig:Comparison_Khan_Rgal}, we present comparisons between the Galactocentric distances reported by \citet{Quireza:06}, \citet{Wenger:19}, and \citet{Khan:24}, and our own determinations. The largest discrepancies are found with the distances from \citet{Quireza:06}, which is not surprising given that those authors used the GRM of \citet{Clemens:85} and \citet{Brand:93}, both of which have been updated and improved in more recent works such as \citet{Reid:14b}. Resulting distance values are presented in Table~2 from the supplentary data. These systematic differences in kinematic distance estimates based on the older GRMs of \citet{Clemens:85} and \citet{Brand:93} have important implications, as they previously introduced biases in the chemical abundance gradients of the inner Galaxy, leading to an apparent flattening of the gradient \citep{Esteban:18}. This artificial flattening is no longer present in more recent studies such as those by \citet{ArellanoCordova:20, ArellanoCordova:21} and \citet{MendezDelgado:22}, which adopted distances are based on Gaia parallaxes \citep{MendezDelgado:20, MendezDelgado:22}.

In Figures \ref{fig:Comparison_Wenger_Rgal} and \ref{fig:Comparison_Khan_Rgal}, we find good consistency between the Galactocentric distances derived with our methodology and those reported by previous authors, with the exception of a slight offset between the distances of \citet{Wenger:19} and ours. This offset is primarily due to the different adopted solar distances, 8.34 kpc in their case and 8.2 kpc in ours. \citet{Khan:24} did not report uncertainties associated with their distances, whereas in our methodology these uncertainties were explicitly determined and propagated. However, the central values are highly consistent with our results. 

\section{Radial Metallicity Gradients in the Milky Way}
\label{Sec:gradients}

Various objects can be used to trace the present-day chemical composition of the ISM in our Galaxy, and in principle, the metallicity distribution derived from each of them should be equivalent. As established in Section~\ref{Sec:Intro}, such tracers include massive stars -- whose lifespans are short enough that their chemical composition, particularly in elements like oxygen, remains largely unaltered -- and the ionised gas in star-forming regions. Since the methodologies used to determine chemical abundances in ionised gas, O-type stars, B-type stars, and Cepheid variables are completely different, comparing the abundance gradients derived from each type of object allows us to identify systematic biases, methodological issues, or specific physical processes affecting each tracer.

\subsection{Metallicity gradient traced by stars} 
\label{subsec:massive_stars_gradient}

%\cesar{He reescrito toda la subsección} 

\begin{figure}
    \includegraphics[width=1.\columnwidth]{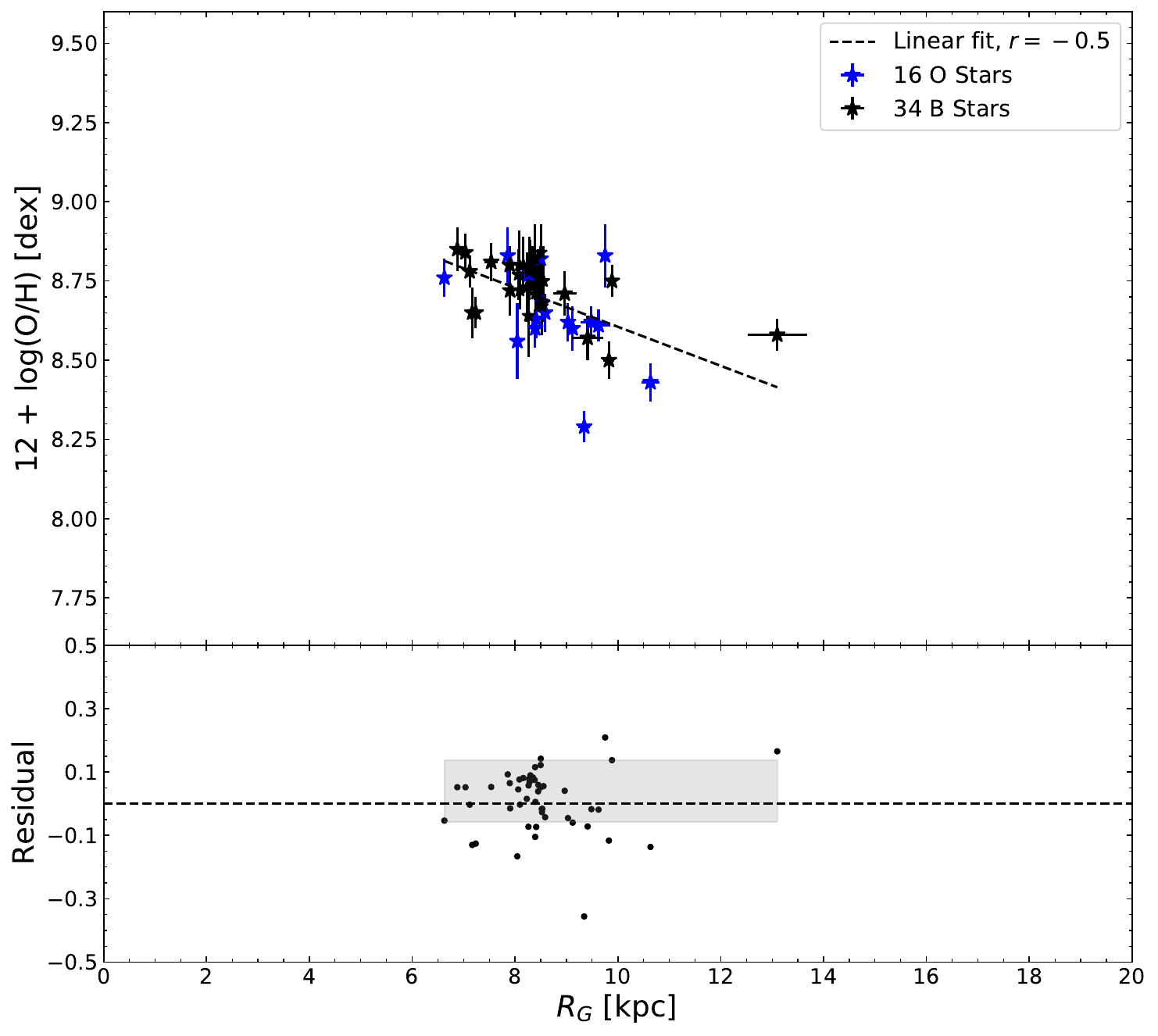}
    \caption{Radial distribution of the O/H ratios in the Milky Way traced by O- and B-type stars. The lower panel displays the residuals and the shaded gray region indicates the typical fitting uncertainty $1\sigma$, corresponding to +0.14 and $-$0.06.}
    \label{fig:OB_gradient}
\end{figure}

In Fig.~\ref{fig:OB_gradient}, we present the O/H gradient derived from the compilation of data for 50 O- and B-type stars from the literature (see Section~\ref{Sec:Observational_Sample} for details). The linear ODR fit is presented in equation~\eqref{eq:gradient_OB}:

\begin{equation}
    \label{eq:gradient_OB}
            12+\text{log(O/H)}_{\text{OB stars}}= -\left(0.06  \pm 0.01 \right) \times R_G + \left(9.2  \pm 0.1\right) \text{ [dex]}.
\end{equation}

Determining a valid gradient from these data is problematic because their radial distribution is quite inhomogeneous; most of them are located in the immediate vicinity of the Sun, and, furthermore, only two of them lie outside the $R_G$ range of 6-10 kpc. Therefore, it does not seem appropriate to assume this gradient to the entire Galactic disk. However, it is worth noting that at the solar Galactocentric distance ($\sim$8.2 kpc), a well-sampled region, these stars are consistent with an abundance of 12+log(O/H)$\approx$ 8.71, essentially the same as the solar abundance reported by \citet{Asplund:21}, 12+log(O/H)$\approx$ 8.69.

\begin{figure}
    \includegraphics[width=1.\columnwidth]{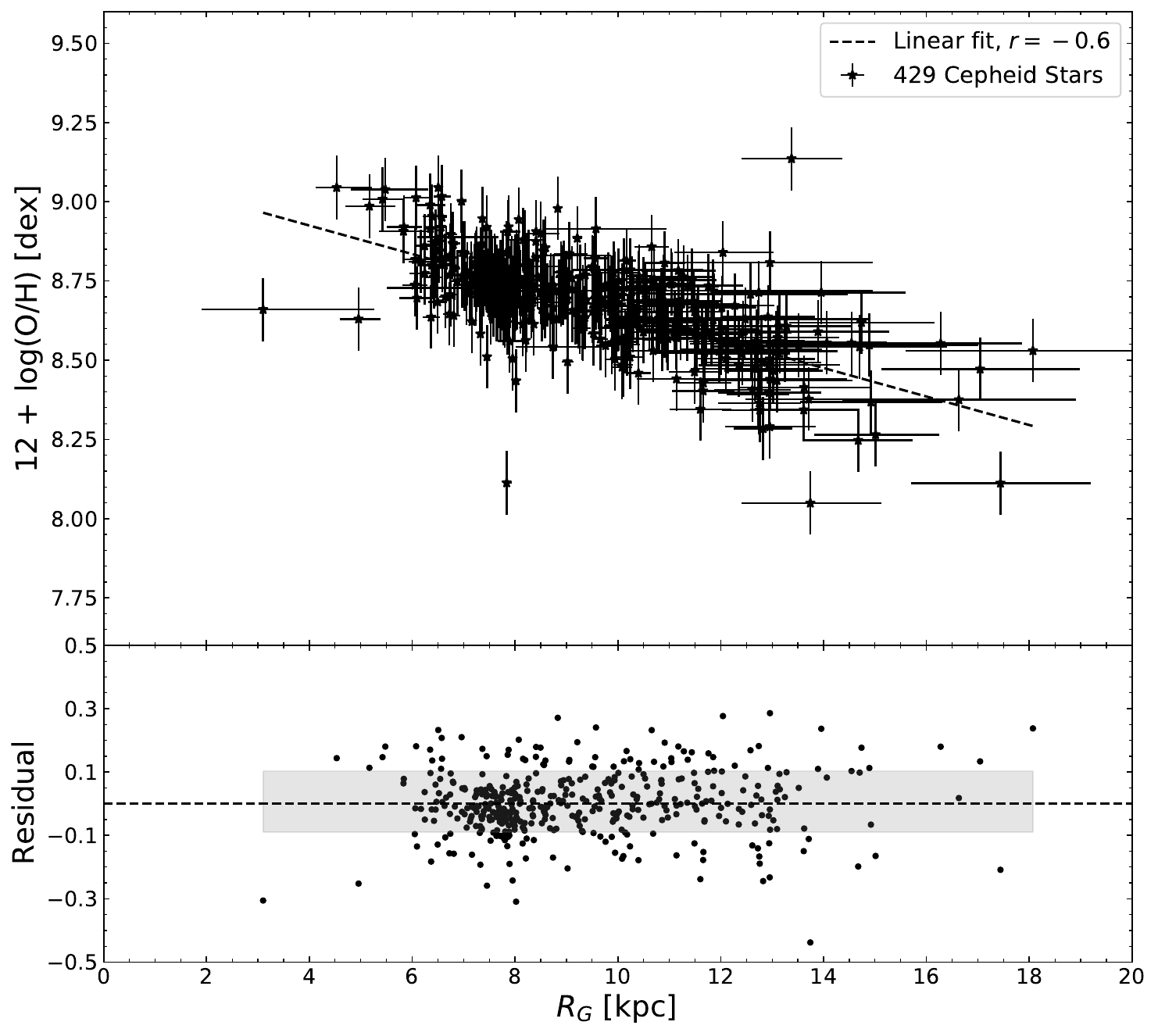}
    \caption{Radial distribution of O/H abundance in the Milky Way traced by Cepheid variable stars. The lower panel displays the residuals and the shaded gray region indicates the typical fitting uncertainty $1\sigma$, corresponding to +0.1 and $-$0.1.}
    \label{fig:cefeidas_gradient}
\end{figure}

In Fig.~\ref{fig:cefeidas_gradient}, we present the O/H abundance gradient derived from 429 classical Cepheid variable stars. These stars, due to their intermediate-to-high masses (typically 3–15 M$_\odot$) and young ages ($\sim$10–300 Myr), are valuable tracers of the present-day chemical composition of the ISM \citep{Lemasle:07,Luck:08}. Importantly, Cepheids are expected to retain the original composition of the ISM from which they formed, especially for elements like oxygen, iron, and the $\alpha$-elements \citep{Luck:18}. The ODR linear fit from Fig. \ref{fig:cefeidas_gradient} is presented in equation~\eqref{eq:gradient_cefeidas}:

\begin{equation}
\label{eq:gradient_cefeidas}
12+\text{log(O/H)}_{\text{Cepheids}}= -\left(0.045 \pm 0.002 \right) \times R_G + \left(9.10 \pm 0.02\right) \text{ [dex]}.
\end{equation}

Comparing figures~\ref{fig:OB_gradient} and \ref{fig:cefeidas_gradient} and the very different uncertainties in the fitting parameters -- a factor of $\sim$7 smaller in the case of Cepheids, the radial gradient defined by Cepheids is much better defined than that of O- and B-type stars. The number of objects is almost an order of magnitude larger for Cepheids and they are much better distributed spatially, spanning a wider distance range.

\subsection{Metallicity gradient traced by \hii~regions}
\label{subsec:hii_gradient}

\begin{figure}
    \includegraphics[width=1.\columnwidth]{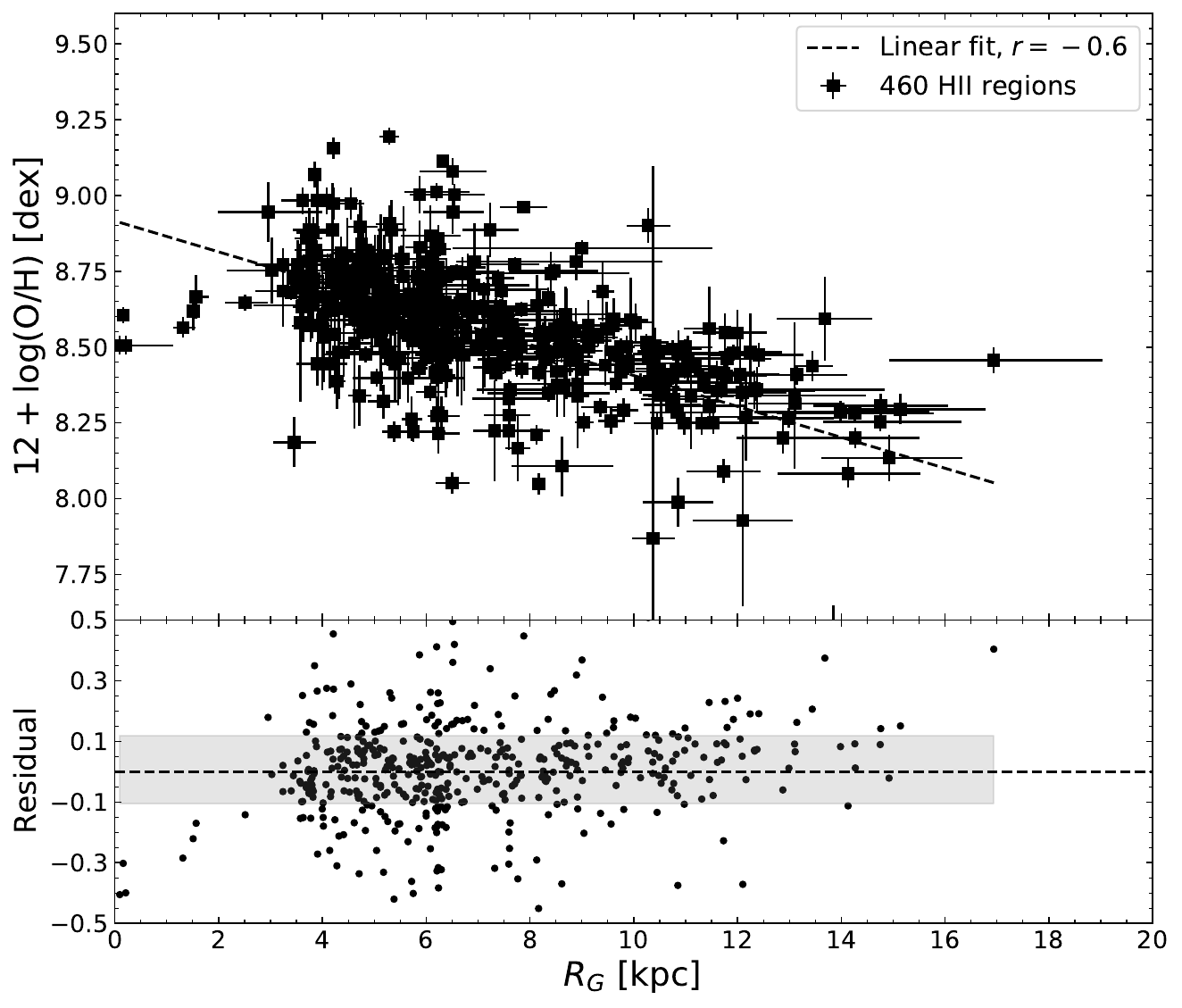}
    \caption{Radial distribution of the O/H ratios in the Milky Way traced by \hii~regions assuming a homogeneous $T_{\rm e}$ structure ($t^2 = 0$). The lower panel displays the residuals and the shaded gray region indicates the typical fitting uncertainty $1\sigma$, corresponding to +0.1 and $-$0.1}
    \label{fig:hii_t2eq0_gradient}
\end{figure}

\begin{figure}
    \includegraphics[width=1.\columnwidth]{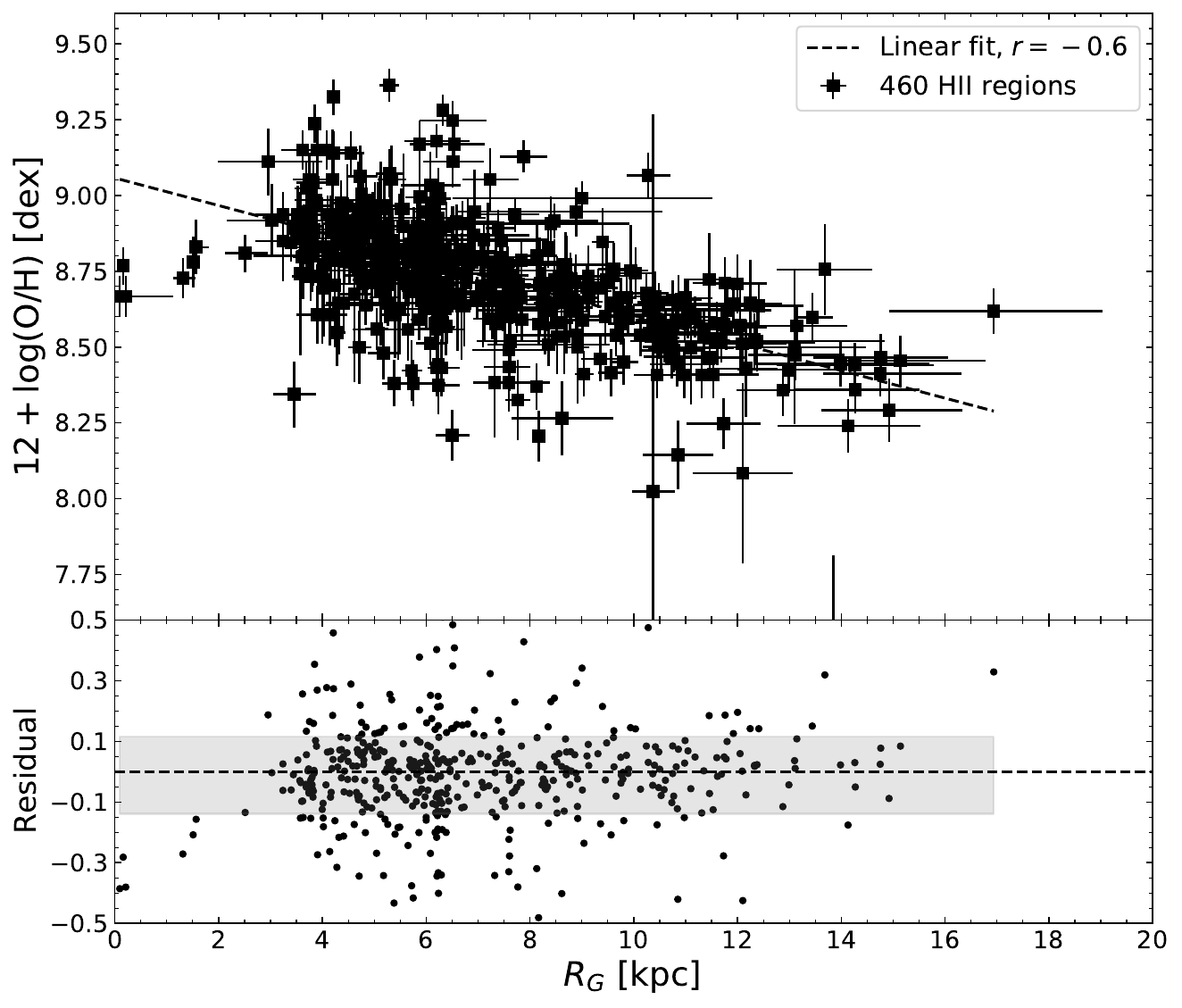}
    \caption{Same as in Fig.~\ref{fig:hii_t2eq0_gradient}, but considering internal temperature variations ($t^2 > 0$). The lower panel displays the residuals and the shaded gray region indicates the typical fitting uncertainty $1\sigma$, corresponding to $+0.1$ and $-0.1$}
    \label{fig:hii_t2geq0_gradient}
\end{figure}

As mentioned in Section~\ref{Subsec:det_tem_metal}, the global metallicity of Galactic \hii~regions can be determined using radio observations that allow for a global measurement of $T_{\rm e}$ in the ionised gas via the radio continuum, which is dominated by hydrogen ions ($T_{\rm e}$(H$^{+}$)). This requires a $T_{\rm e}$–metallicity relation such as the ones we have derived in equations~\eqref{eq:tem_metal_t2eq0} and \eqref{eq:tem_metal_t2gt0}, corresponding to the cases without and with internal $T_{\rm e}$ inhomogeneities, i.e., $t^2 = 0$ and $t^2 > 0$, respectively. The results, based on the 460 \hii~regions observed in radio, are shown in Figs.~\ref{fig:hii_t2eq0_gradient} and \ref{fig:hii_t2geq0_gradient} for each case, respectively. In both cases, the sample of \hii~regions satisfactorily covers the range of Galactocentric distances between 3 and 17 kpc.

The ODR fits are presented below for the $t^2 = 0$ case:

\begin{equation}
\label{eq:gradient_t2zero}
12+\text{log(O/H)}_{t^2=0}= -\left(0.051 \pm 0.003\right) \times R_G + \left(8.92 \pm  0.02\right) \text{ [dex]},
\end{equation}

and for the $t^2 > 0$ case:

\begin{equation}
\label{eq:gradient_t2notzero}
12+\text{log(O/H)}_{t^2>0}= -\left(0.045 \pm 0.003\right) \times R_G + \left(9.06 \pm  0.02\right) \text{ [dex]}.
\end{equation}

It is remarkable that both radial gradient determinations are very consistent with the slope obtained in the most recent and precise calculations of the Galactic abundance gradient based on direct determination of oxygen abundances in \hii~region spectra obtained with the 10.4m GTC telescope \citep{ArellanoCordova:20, ArellanoCordova:21,MendezDelgado:22}. \citet{MendezDelgado:22} present the latest improved radial gradient parameters using distances derived from Gaia EDR3 parallaxes, finding a slope of $-0.044\pm0.009$ for the $t^2 = 0$ case and $-0.059\pm0.012$ for $t^2 > 0$. In this latter $t^2>0$ case, however, the methodology of \citet{MendezDelgado:22} differs slightly from ours, since these authors considered the presence of temperature inhomogeneities in both the high- and low-ionisation zones, whereas in the present work they were taken into account only in the high-ionisation zone, following the prescriptions by \citet{MendezDelgado:23a}.

Typically, studies of the Galactic metallicity distribution using radio observations have relied on the $T_{\rm e}$–metallicity relation from \citet{Shaver:83}, presented in their equation~(16). However, this relation has significant limitations: it is based on outdated atomic data -- which form the foundation for determinations of physical conditions and ionic abundances -- and on optical observations obtained before the advent of CCD technology. Observations of Galactic \hii~regions have since been systematically improved, particularly through the use of 8–10 m telescopes with long exposure times \citep{Esteban:04,Esteban:17,Esteban:18,GarciaRojas:04,GarciaRojas:05,GarciaRojas:06,GarciaRojas:07,MesaDelgado:09}. These technological advances have allowed us to build our adopted DESIRED-E sample, which includes 225 \hii~regions with simultaneous detections of both $T_{\rm e}$(\nii) and $T_{\rm e}$(\oiii) --a condition achieved in only three regions by \citet{Shaver:83} in their sample. This limitation in earlier work forced the use of additional calibrations based on photoionisation models to connect the $T_{\rm e}$ of the low- and high-ionisation zones, thereby undermining the ``direct'' nature of the so-called ``direct method''. Furthermore, as discussed in Section~\ref{Subsec:resulting_temperature_relations}, the range of $T_{\rm e}$ values covered by \citet{Shaver:83} is comparatively narrower, spanning only from $\sim 6000$ K to $\sim 11000$ K. However, although the $T_{\rm e}$–metallicity relation of \citet{Shaver:83} was calibrated over a narrow temperature range, its use outside this interval is widespread \citep{Quinet:96,Wenger:17,Wenger:19} and constitutes a mathematical extrapolation that is not supported by observational data.

More recently, \citet{Balser:24} analysed the $T_{\rm e}$–metallicity relation using photoionisation models with the \textit{Cloudy} code \citep{Ferland:17}, finding good agreement with the relation of \citet{Shaver:83}. This is not surprising given the dispersion observed across different model combinations. In that sense, the empirical relations derived in this work (equations~\eqref{eq:tem_metal_t2eq0} and \eqref{eq:tem_metal_t2gt0}) would also fall within the broad family of relations found by \citet{Balser:24}. A major limitation of photoionisation models is the difficulty in constraining which models are physically realistic and which represent conditions unlikely to be found in nature. This can artificially broaden the range of possible $T_{\rm e}$-metallicity relations that can be modelled. In contrast, our relations (equations~\eqref{eq:tem_metal_t2eq0} and \eqref{eq:tem_metal_t2gt0}) are derived exclusively from observations, and are therefore firmly realistic.

\begin{figure}
    \includegraphics[width=1.\columnwidth]{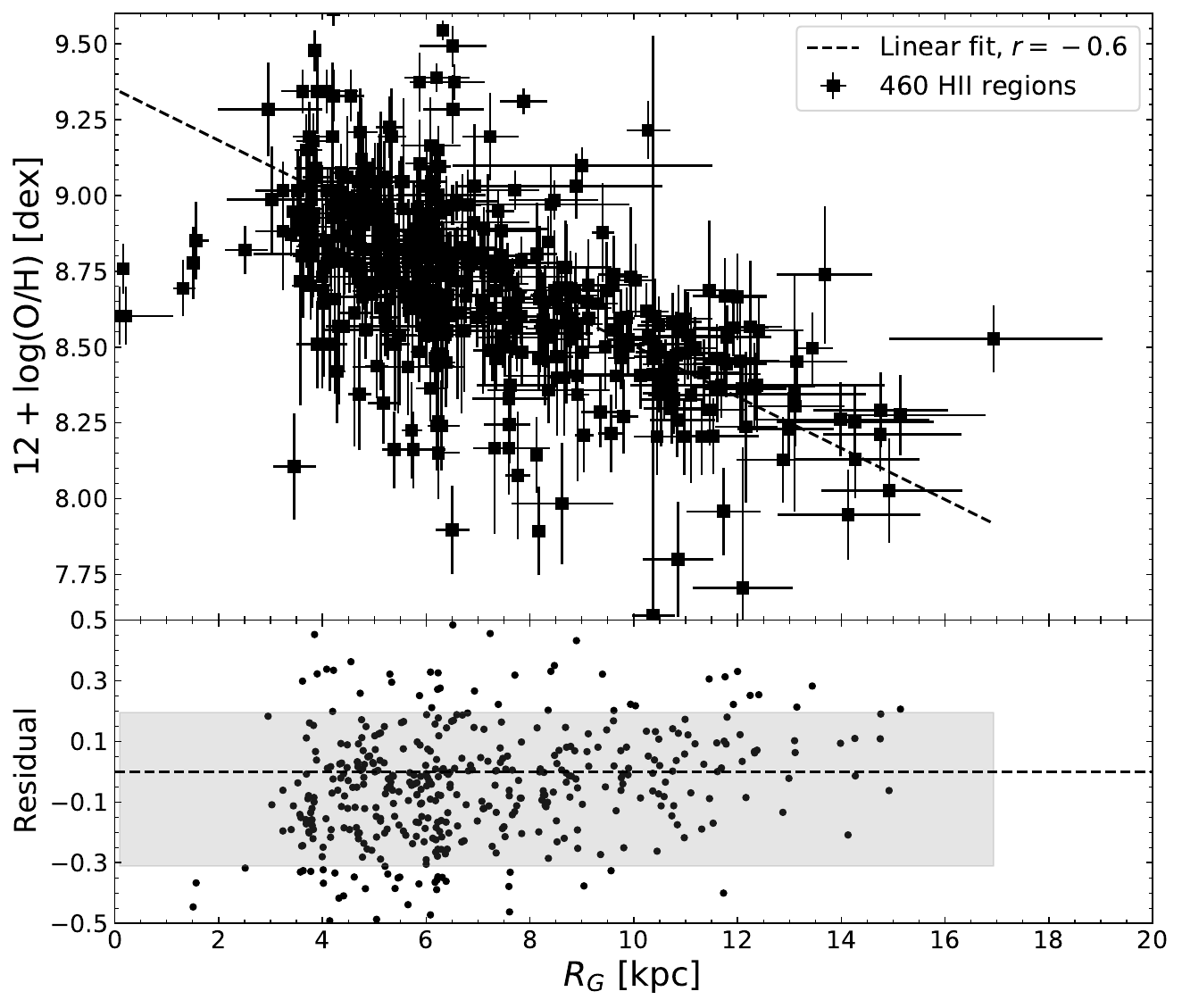}
    \caption{Radial distribution of the O/H ratios in the Milky Way traced by \hii~regions assuming the $T_{\rm e}$-metallicity relation of \citet{Shaver:83}. The lower panel displays the residuals. The shaded gray region indicates the typical fitting uncertainty $1\sigma$, corresponding to +0.2 and $-$0.3.}
    \label{fig:hii_shaver_gradient}
\end{figure}

In any case, given the widespread use of the $T_{\rm e}$–metallicity relation determined by \citet{Shaver:83}, we have applied their equation~(16) to our sample of \hii~regions observed in radio and derived the corresponding O/H gradient. The result is presented in Fig.~\ref{fig:hii_shaver_gradient} and equation~\eqref{eq:gradient_shaver}:

\begin{equation}
\label{eq:gradient_shaver}
12+\text{log(O/H)}_\text{Shaver+83}= -\left(0.085 \pm 0.005\right) \times R_G + \left(9.35 \pm 0.04\right) \text{ [dex]}.
\end{equation}

\section{Nebular versus Stellar Radial Metallicity Gradients in the Milky Way}
\label{sec:nebular_vs_stellar}

\begin{figure*}
    \includegraphics[width=2.\columnwidth]{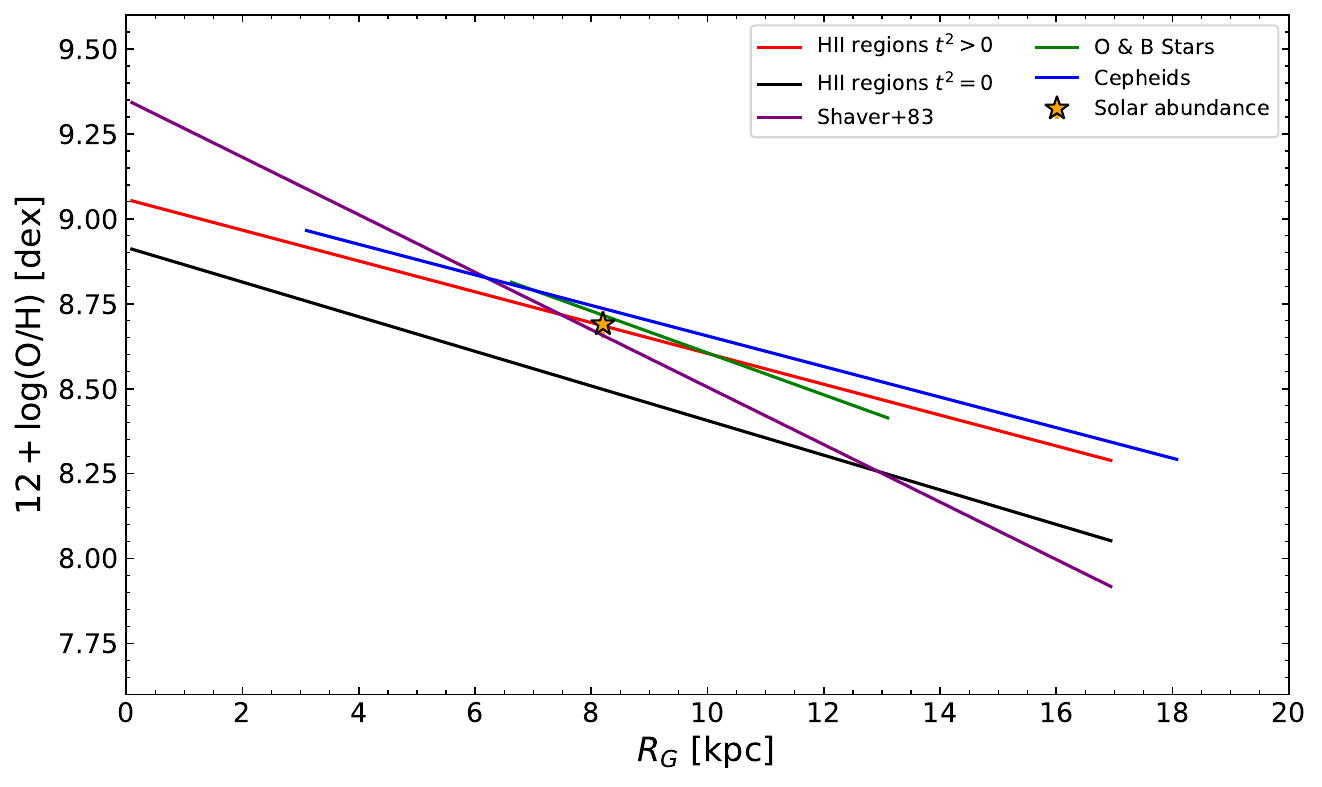}
    \caption{Galactic metallicity gradients obtained using various tracers of the current ISM composition. In the case of \hii~regions observed in radio, three different approaches were explored, each with different physical implications: assuming a homogeneous thermal structure ($t^2 = 0$), considering the presence of temperature variations ($t^2 > 0$), and using the classical calibration from \citet{Shaver:83}, which also assumes $t^2 = 0$ but was based on pre-CCD optical observations and outdated atomic data. As a reference, the solar O/H value determined by \citet{Asplund:21} is indicated with an orange star. The uncertainties associated with the gradients are not shown here to allow for a clearer visual comparison, but they can be found individually in Figs.~\ref{fig:OB_gradient}-\ref{fig:hii_shaver_gradient}.}
    \label{fig:different_relations}
\end{figure*}

In Fig.~\ref{fig:different_relations}, we present a comparison of the different metallicity gradients obtained in this study, using both stellar objects and \hii~regions observed in radio. The corresponding relations are given in equations~\eqref{eq:gradient_OB}, \eqref{eq:gradient_cefeidas}, \eqref{eq:gradient_t2zero}, \eqref{eq:gradient_t2notzero}, and \eqref{eq:gradient_shaver}. Notably, there is good agreement --both in slope and intercept-- between the nebular gradients derived under the $t^2 > 0$ assumption and the stellar gradients obtained from O- and B-type and classical Cepheid stars. In the case of O- and B-type stars, the slope is slightly steeper than those found for \hii~regions and Cepheids, but this may simply reflect their limited range of Galactocentric distances.

In contrast, the gradient obtained from nebular regions under the ``direct method'' assumption ($t^2 = 0$) shows an offset of up to 0.3 dex when compared to the other tracers. This discrepancy is difficult to justify, as the abundances of young stars are not expected to have increased in oxygen during their lifetimes. In fact, if CNO-cycle products were present in the photosphere of these stars, the overall effect would be to reduce the oxygen abundance in favour of nitrogen. Some authors have suggested that the oxygen depletion onto dust grains could help mitigate this discrepancy \citep{Bresolin:16, Bresolin:25}. However, the observed differences are far too large to be explained by this effect. Most studies of element depletion in dust are based on neutral nebular environments \citep{Jenkins:09}, whose physical conditions differ substantially from those of ionised gas. Even in those neutral regions, oxygen belongs to the group of volatile elements, with observed depletions typically around 0.2 dex \citep{Jenkins:09, Ritchey:23, Konstantopoulou:24}.

It is worth noting that the precise molecular carriers responsible for oxygen depletion remain uncertain, though a significant fraction is expected to come from molecules like O$_2$, which cannot survive in ionised gas due to the intense radiation fields from massive ionising stars \citep{Osterbrock:06, Jenkins:09}. Studies of dust within ionised gas have revealed an upper limit of $\sim$0.1 dex for oxygen depletion, as observed in the Orion Nebula, which is expected to have near-solar metallicity \citep{MesaDelgado:09}. Studies of Fe depletion onto dust grains in ionised environments have shown a strong gradient, with higher depletion in regions of higher metallicity \citep{Rodriguez:05, Izotov:06, MendezDelgado:24b}. Assuming a similar trend for oxygen, the fraction of oxygen locked in dust would be negligible in sub-solar metallicity regions, considering the upper limit of $\sim$0.1 dex found in Orion \citep{MendezDelgado:24b}.

The above discussion clearly indicates that the discrepancy between stellar and nebular abundances under the assumption of a homogeneous $T_{\rm e}$ structure cannot be fully attributed to dust depletion effects. In Fig.~\ref{fig:different_relations}, we show that the discrepancy between stellar and $t^2 = 0$ nebular abundances is an observationally robust result for, at least, metallicities ranging from 12 + log(O/H) $\sim$ 8.4 to $\sim$ 9. An interesting potential exception is the case of M20. Using integral field spectroscopic observations from the LVM, \citet{Sattler:26} find good agreement between O/H abundances derived from CELs assuming $t^2=0$ and those determined for the ionizing star HD 164492 \citep{Martins:15}. However, the uncertainties in the stellar abundances are as large as 0.3 dex, which leaves the discrepancy between abundance methods as an open issue for this nebula.

In works such as \citet{Bresolin:16} and \citet{Bresolin:22}, it has been suggested that this stellar and nebular discrepancy only appears at O/H values higher or equal than solar 12 + log(O/H) $\sim$ 8.7 \citep{Asplund:21}, while at lower abundances, massive stars and \hii~regions analysed via the ``direct method'' yield consistent results, thereby reinforcing the validity of the latter. However, as pointed out by \citet{MendezDelgado:24b}, most of the metallicity estimates adopted or derived by \citet{Bresolin:16, Bresolin:22, Bresolin:25} are actually based on iron-peak elements and then converted to O/H using the solar Fe/O ratio. This assumption is not generally valid, as the solar Fe/O ratio is not universal and is known to vary by more than an order of magnitude depending on the evolutionary state of the galaxy \citep{Amarsi:19, Chruslinska:23}.

Regarding the direct determinations of the O/H ratio in extragalactic stellar systems adopted by \citet{Bresolin:16}, \citet{MendezDelgado:24b} emphasised that many of those stars show N/H overabundances of more than an order of magnitude, indicating strong mixing due to CNO processing. This can lower the observed oxygen abundance, modifying it from its initial value at the time of the star's formation. Unfortunately, the narrative of ``good consistency'' between massive stars and nebular determinations based on the ``direct method'' has been echoed in various review studies \citep{Maiolino:19, Curti:25}. It is therefore important to emphasise that this agreement does not hold --at least not from stellar metallicities as low as 12 + log(O/H) $\sim$ 8.4.

The gradient obtained using the $T_{\rm e}$–metallicity relation from \citet{Shaver:83} presents its own challenges, despite being derived under the $t^2 = 0$ assumption. The resulting gradient is significantly steeper than those obtained through other methods, leading to a considerable overestimation of the O/H abundances in the inner Galaxy ($R_G < 8$ kpc), while at Galactocentric distances beyond 12 kpc it tends to underestimate the metallicity --even more so than the gradient derived here using the DESIRED sample under the $t^2 = 0$ assumption. The discrepancies are particularly striking: at $R_G \sim 14$ kpc --a range well covered by radio observations-- the gradient obtained using the \citet{Shaver:83} $T_{\rm e}$-metallicity relation underestimates the metallicity by up to 0.5 dex compared to values inferred from Cepheid stars or \hii~regions analysed assuming $t^2 > 0$. This mismatch arises from the various caveats discussed in Section~\ref{subsec:hii_gradient}. While the foundational contribution of \citet{Shaver:83} to the field must be acknowledged, it is equally important to appreciate the significant theoretical and observational advancements made in recent years, which now allow for a more accurate determination of the $T_{\rm e}$–metallicity relation for \hii~regions observed in the radio. It is important to emphasize that the temperature range used to calibrate the relation presented by \citet{Shaver:83} is quite narrow; nevertheless, it is typically extrapolated beyond this interval, yielding results consistent with those shown in our Fig.~\ref{fig:different_relations}.

As clearly shown in Fig.~\ref{fig:metal_tem_t2geq0}, although incorporating $T_{\rm e}$ inhomogeneities naturally resolves the discrepancy between stellar and nebular abundances, the scatter in the $T_{\rm e}$–metallicity relation under the $t^2 > 0$ assumption is $\sim$0.1 dex larger than that obtained assuming $t^2 = 0$. This increased scatter is a direct result of the larger typical observational uncertainties associated with $T_{\rm e}$(\nii), which forms the basis of the $t^2 > 0$ estimation, compared to those of $T_{\rm e}$(\oiii). In fact, the study of temperature relations within the DESIRED sample (Orte-García et al., in prep.) reveals a significant correlation between the uncertainties in $T_{\rm e}$(\nii) and its central value. This behaviour is expected, as metal-poor regions tend to exhibit higher ionisation states \citep{Peimbert:74,Pagel:92}, leading to an extremely low N$^+$/N ratio, which makes the detection of \nii~$\lambda 5755$ particularly challenging. Ongoing observations within the DESIRED project are specifically targeting the simultaneous measurement of both $T_{\rm e}$(\nii) and $T_{\rm e}$(\oiii), given the significant implications this has for the high-redshift Universe revealed by the JWST. To date, there are no reliable $T_{\rm e}$ determinations in the low-ionisation zone at high redshift, with only a few objects having $T_{\rm e}$(\oii) measurements \citep{Sanders:23}, which often differ from $T_{\rm e}$(\nii) \citep{Rogers:21, Zurita:21,MendezDelgado:23b, RickardsVaught:24} due, in part, to its greater dependence on the assumed value of  $n_{\rm e}$ \citep{MendezDelgado:23b}.

\subsection{The Radial Oxygen Gradient in the Context of Galactic Chemical Evolution}
\label{Sec:theor}

Our observational findings have important implications for Galactic chemical evolution models (CEMs), which aim to reproduce the spatial distribution of chemical abundances observed in stars of different ages and in the ISM. It is not our purpose to provide a detailed chemical evolution model here, leaving such work to future studies, but it is worthwhile to offer a brief comment on this aspect. In these models, the negative radial oxygen abundance gradient arises naturally from an inside-out formation scenario of the Milky Way disc, in which the inner regions experience earlier and more rapid chemical enrichment than the outer ones \citep{Carigi:19}. In addition, the gradual dilution of the ISM by the infall of metal-poor gas (as well as radial mixing processes) is commonly invoked to moderate the steepening of the gradient over time, helping to reproduce the relatively shallow slope (approximately $-0.04$ to $-0.05,\mathrm{dex\ kpc^{-1}}$) observed in the present-day Galaxy.

Our results --particularly the agreement between the nebular O/H gradient (under the $t^2>0$ assumption) and that traced by young stars-- suggest that no extraordinary processes beyond the standard ingredients of CEMs are required. In particular, there is no clear evidence demanding a strong depletion of O onto dust, nor invoking ISM dilution occurring between the epoch of star formation and the present time \citep{Spitoni:23}. Instead, a possible metal-rich and relatively low-mass accretion event remains viable, since such an infall would not dilute the ISM oxygen abundance and could have triggered, in recent times, the solar-metallicity star formation rate inferred by \citet{FernadezAlvar:25}.

Furthermore, the absolute value of the present-day O/H gradient provides insight into the relative number of massive stars formed in each star-formation episode, because oxygen is predominantly produced by stars in the $8$--$M_{\mathrm{up}},M_{\odot}$ range, where $M_{\mathrm{up}}$ denotes the upper mass limit of the IMF. The fraction of massive stars in a galaxy is primarily determined by $M_{\mathrm{up}}$, assuming a fixed IMF slope in the high-mass regime. Under these assumptions, \citet{Carigi:19} inferred $M_{\mathrm{up}} = 40 M_{\odot}$ (or $80 M_{\odot}$) when their Galactic disc CEMs were constrained by O/H values from \ion{H}{ii} regions derived with $t^{2}=0$ (or $t^{2}>0$), respectively. From the O/H gradient obtained in this paper for $t^{2}>0$ (see Eq.~\eqref{eq:gradient_t2notzero}), we estimate $M_{\mathrm{up}} \approx 60 M_{\odot}$.

\section{Azimuthal Metallicity distribution in the Milky Way}
\label{Sec:Azimuthal}

Galaxies exhibit pronounced radial metallicity gradients that trace their inside-out enrichment history, but it remains unclear whether the interstellar medium is chemically homogeneous at a given radius or if significant azimuthal abundance variations occur \citep{DeCia:21,Esteban:22, Ritchey:23}. Any such azimuthal metallicity differences would provide valuable insight into the efficiency of mixing processes \citep{Kreckel:19, Kreckel:20,Otto:25}. Dynamic features --such as spiral arms-- can induce gas flows and localised enrichment or dilution, potentially leading to non-uniform abundances along a ring \citep{Hawkins:23}. These inhomogeneities matter because metals regulate ISM cooling and star formation; thus, variations in metallicity could translate into local differences in star-forming conditions \citep{SanchezMenguiano:16,Ho:17}.

Various models of galaxy formation and evolution predict the existence of azimuthal metallicity variations, either as a consequence of peculiar motions near spiral arms or as the result of density fluctuations that drive stellar migration and the redistribution of metals \citep{DiMatteo:13,Grand:16,Spitoni:19}. However, the predicted amplitude of such variations is typically of the order of $\sim 0.1$ dex or less. This poses a challenge for their reliable detection with nebular objects, since chemical abundance determinations generally involve uncertainties of the same order, or even larger. In some cases, the availability of multiple high-quality spectra can reduce the impact of statistical errors well below this threshold. Nevertheless, the presence and quantification of systematic uncertainties arising from factors such as temperature inhomogeneities ($t^2 > 0$) or the depletion of certain elements onto dust grains remains problematic, as demonstrated in this work.

\begin{figure*}
    \includegraphics[width = 2.\columnwidth]{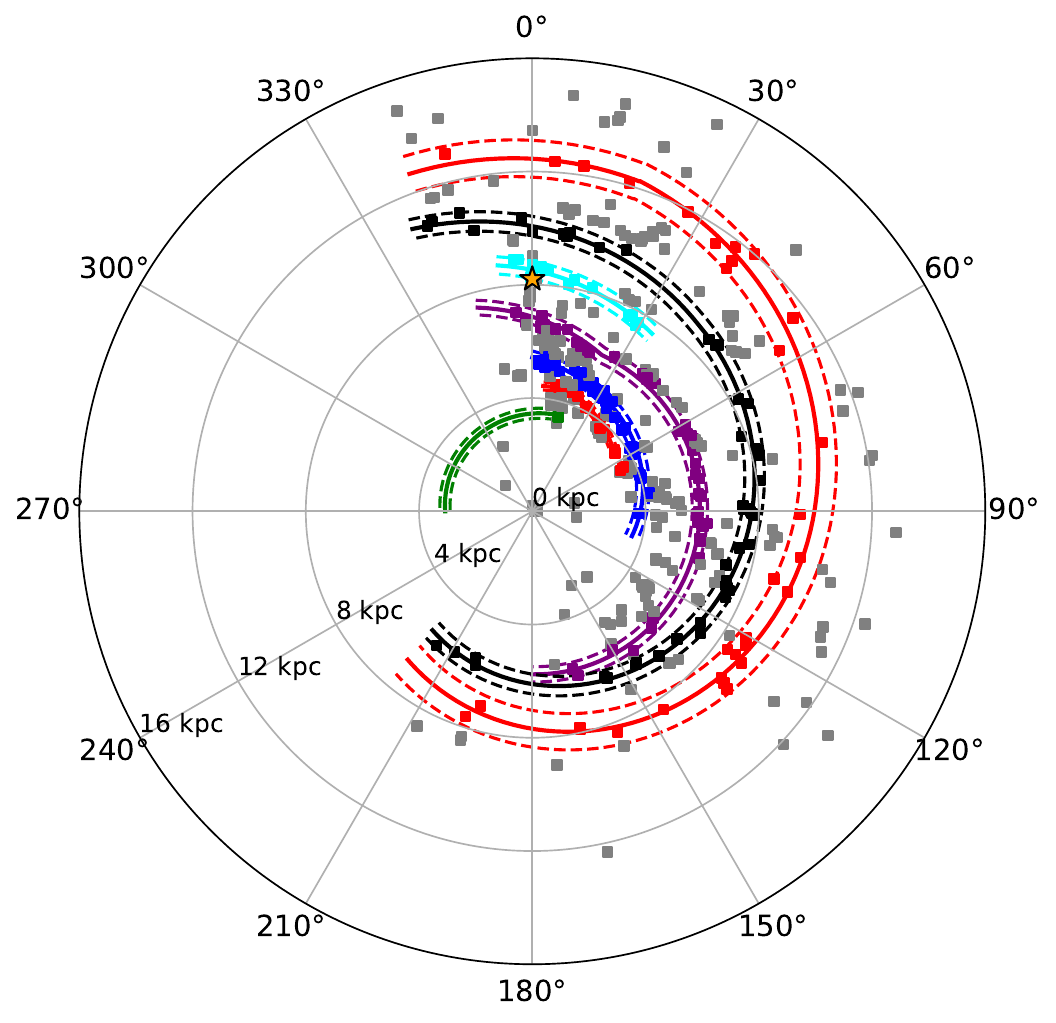}
    \caption{Face-on Galaxy projection in polar coordinates $(R_G, \beta)$, the squares represents the 460 \hii~regions observed in radio and the assignment of these \hii~regions to spiral arms from \citep{Reid:19} is discussed in the section \ref{Sec:Azimuthal}, the solid curved lines trace the centres of the spiral arms and the dashed line represents the $1\sigma$ widths. We extrapolate the spiral arms to cover our sample range in galactic azimuth $\beta$. The assigned colours are: 3kpc arm, green; Norma-Outer arm, red; Scutum–Centaurus–OSC arm, blue; Sagittarius–Carina arm, purple; Local arm, cyan; Perseus arm, black. The sun (orange star) is at (0, 8.2) and the grey squares indicate sources for which the arm assignment is unclear.}
    \label{fig:azimuthal_t2eq0}
\end{figure*}

In Fig.~\ref{fig:azimuthal_t2eq0}, we show the distribution of the 460 \hii~regions in our sample observed in radio wavelengths, displaying a broad azimuthal coverage of at least half of the Galaxy. The figure also includes the spiral arm loci from \citet{Reid:19}, as well as the position of the Sun, marked with an orange star.

\begin{figure*}
    \includegraphics[width=2.\columnwidth, height = 0.3\linewidth]{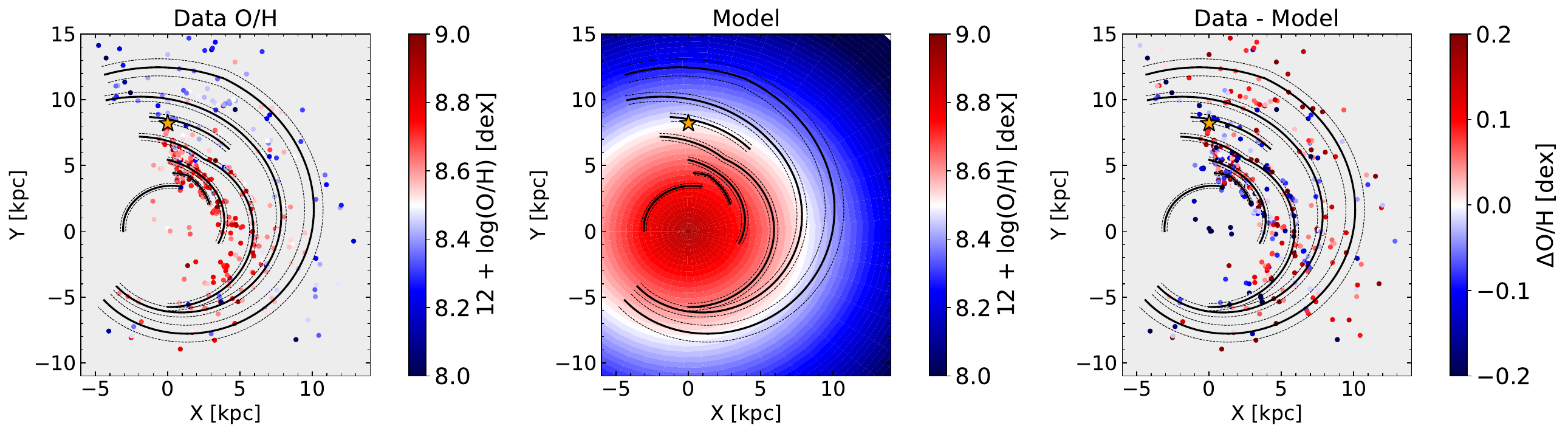}
    \caption{Left Panel: Face-on distribution for the 460 \hii~regions observed in radio indicating the O/H abundance derived by the Temperature-Metallicity relation describe by equation~\eqref{eq:tem_metal_t2eq0} which assume a homogeneous temperature structure ($t^2=0$). Middle Panel: Face-on projection of radial metallicity gradient defined by the equation~\eqref{eq:gradient_t2zero} which is the our best fitting model when we assume an homogeneous temperature structure ($t^2=0$). Right Panel: Face-on distribution of the residuals in the O/H abundance relative to the radial metallicity gradient when we assume an homogeneous temperature structure ($t^2=0$). In all the panels the spiral arms with their widths from \citep{Reid:19} are shown as black lines and the position of the sun from \citep{Bland:16} as an orange star.  
    }
    \label{fig:Face-On_t2eq0}
\end{figure*}

To investigate the presence (or absence) of evident azimuthal metallicity inhomogeneities, we focus on the case $t^2=0$, which yields the lowest statistical noise. For the case $t^2>0$,the results remain equally consistent. We use the O/H distribution shown in Fig.~\ref{fig:hii_t2eq0_gradient}, subtract the radial gradient, and plot the residuals in face-on projection maps. In Fig.~\ref{fig:Face-On_t2eq0}, the left panel shows the distribution of the Galactic \hii~regions, with a colour bar indicating the inferred O/H abundance; the central panel displays the radial gradient in O/H abundances, as described by Eq.~\eqref{eq:gradient_t2zero}; and the right panel presents the residuals obtained after subtracting the radial gradient from the individual O/H values of the left panel.

\begin{figure}
    \centering
    \includegraphics[width=1\columnwidth]{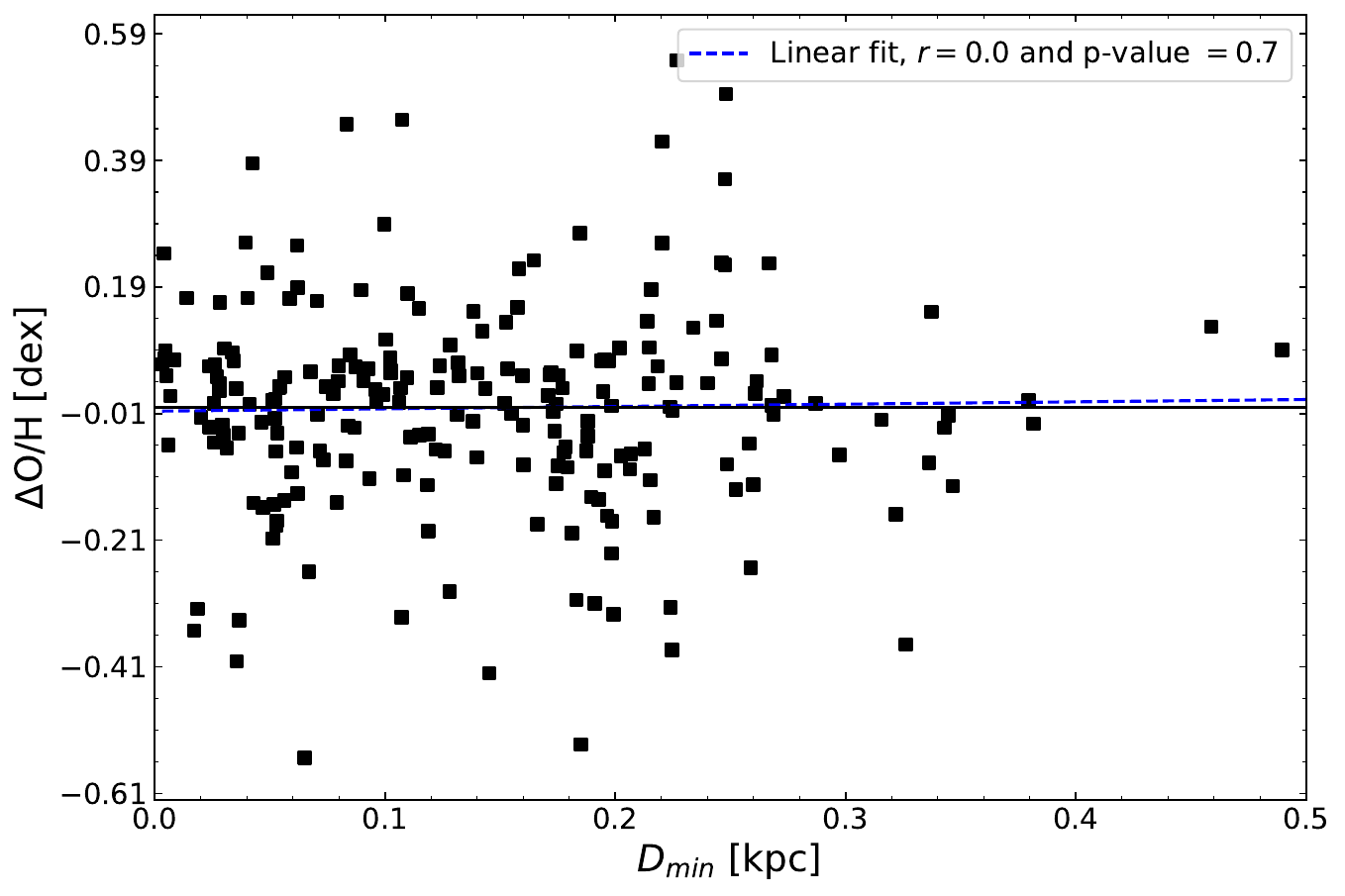}
    \caption{The residuals in O/H abundances relative to the radial metallicity gradient, assuming an homogeneous temperature structure ($t^2 =0$), as a function of the minimum distance ($D_{min}$) between the radio-observed \hii~regions and their respective spiral arms.}
    \label{fig:hii_dmin_t2eq0}
\end{figure}

At first glance, no clear azimuthal trend is observed in the residuals, which follow a statistical distribution with $\sigma \approx 0.1$ dex. This level of uncertainty does not appear to correlate with the spatial location of the spiral arms. To investigate this issue in more detail, we examined the relation between the O/H residuals and the distance to the nearest Galactic arm, considering only regions lying up to $1\sigma$ of the arm’s central position in order to avoid overlap between adjacent arms. The results are shown in Fig.~\ref{fig:hii_dmin_t2eq0}, where no obvious trend can be discerned.

Our results indicate that, if azimuthal metallicity variations driven by the spiral arms exist within the sampled region of the Milky Way, they must be smaller than $\sim 0.1$ dex, consistent with the recent findings of \citet{Kreckel:25} based on $T_{\rm e}$(\nii)-derived metallicities in extragalactic systems. This suggests that the Galactic nebular gas is relatively well mixed, not only in the solar neighbourhood, as previously reported by \citet{Esteban:22}, but also across considerably larger areas such as those covered in this work.

Our interpretation of the azimuthal variations inferred from the metallicity distribution of the radio observed \hii~regions partially differs from that of \citet{Wenger:19}. Although we agree that the apparent metallicity variations are of the order of $\sim 0.1$ dex, we emphasize that the typical uncertainties in nebular metallicity determinations based on radio-$T_{\rm e}$ determinations have the same magnitude. This prevents us from drawing firm conclusions about the presence or absence of azimuthal variations on scales smaller than this level. While it is possible to construct contour maps of the residual distribution, as done by \citet{Wenger:19}, their connection to a clearly defined Galactic structural component is not obvious. In addition, although the idea of determining gradients with intercepts and slopes for different azimuthal subregions is promising, the various azimuthal slices are not uniformly sampled, which may introduce additional systematic uncertainties.

It should be emphasized that, in this work, particular care was taken in the determination of chemical abundances, making use of the highest-quality spectra available in the literature, both in the optical and radio domains. Unfortunately, the precision limits of the nebular determinations mentioned above pose significant observational challenges for their detection and quantification. This is consistent with what has been reported in nearby spiral galaxies \citep{SanchezMenguiano:16,SanchezMenguiano:20, Ho:17, Kreckel:19,Kreckel:20, Kreckel:25}, where azimuthal variations are common but subtle, often exhibiting asymmetries across galaxy disks and no clear systematic correlations with galactic disk environment

In many extragalactic investigations of nebular gas metallicity gradients --particularly those based on surveys designed to provide extensive coverage across large galaxy samples--, the direct determination of chemical abundances is not feasible, as it requires the detection of at least one faint auroral line sensitive to the physical conditions of the gas. As a result, most of these studies rely on statistical calibrations, commonly referred to as ``strong-line methods'' \citep{Pilyugin:16}. Such calibrations carry uncertainties considerably larger than 0.1 dex, owing to their statistical nature. Nevertheless, these uncertainties are often ``hidden'', since they are not always explicitly reported or properly propagated in the analysis, since many of them are systematic and not statistical in origin \citep[see, for example, the excellent discussion in the Appendix of][]{ValeAsari:16}. In such cases, there is a risk of confusing systematic errors inherent to the calibrator with genuine spatial variations in the derived quantities.

We consider it necessary to contrast statistically large yet high-quality samples --such as the quality level employed in the DESIRED projects-- with previous findings based on strong-line calibrations, in order to shed light on the presence of such variations across multiple galactic systems. This approach allows for a fair comparison between other galaxies and the Milky Way, where we find no evidence of azimuthal metallicity variations larger than $\sim$0.1 dex. However, given the difficulties in reliably estimating the level of uncertainty in heavy-element abundance determinations --even when using large, high-quality datasets-- due to the presence of poorly understood systematic errors, we consider it necessary to adopt a cautious approach when interpreting chemical abundance residuals in different galactic environments.

\section{Conclusions}
\label{Sec:conclusions}

As part of the DESIRED-E project \citep{MendezDelgado:23b}, we have analysed a sample of 225 star-forming regions with deep optical spectra and simultaneous detections of $T_{\rm e}$(\nii) and $T_{\rm e}$(\oiii) --the largest such sample available in the literature. Using this unprecedented dataset, we have derived new empirical $T_{\rm e}$–metallicity relations that link the gas-phase oxygen abundance (O/H) to the global $T_{\rm e}$ inferred from radio observations, $T_{\rm e}$(H$^+$). These relations were derived under two scenarios for the thermal structure of the ionised gas: (1) a homogeneous $T_{\rm e}$ distribution ($t^2=0$), corresponding to the commonly used ``direct method'', and (2) an inhomogeneous $T_{\rm e}$ structure ($t^2>0$), incorporating internal $T_{\rm e}$ fluctuations as described in the formalism of \citet{Peimbert:67}. The existence of $T_{\rm e}$ fluctuations has been widely proposed as an explanation for the well-known discrepancy between abundances derived from CELs and those from RLs, with the latter systematically yielding higher values \citep{MendezDelgado:23a}. When adopting $t^2>0$, CEL-based abundances become consistent with those derived from RLs \citep{Esteban:04, GarciaRojas:07b}. The resulting relations are presented in equations~\eqref{eq:tem_metal_t2eq0} and \eqref{eq:tem_metal_t2gt0}.

To test the validity of these relations, we applied them to a sample of 460 Galactic \hii~regions observed in radio by \citet{Quireza:06a}, \citet{Wenger:19}, and \citet{Khan:24}, using their available $T_{\rm e}$(H$^+$) values. We derived the radial metallicity gradients under both the $t^2=0$ and $t^2>0$ assumptions. Accurate Galactocentric distances were computed using either parallaxes of associated stars or compact sources, or, when not available, kinematic distances. These kinematic distances were recalculated in this work using a consistent Galactic rotation curve model from \citet{Reid:14b}, which has been shown to agree with parallax-based distances \citep{MendezDelgado:22}, and implemented via the Monte Carlo approach described by \citet{Wenger:18}, ensuring proper propagation of uncertainties.

To further assess which thermal structure assumption is more physically plausible, we also determined O/H gradients based on 50 young massive B and O-type stars and 429 classical Cepheids under the premise that their surface O/H abundances reflect the original ISM composition at their birth.

We find that the $T_{\rm e}$–metallicity relation incorporating $T_{\rm e}$ fluctuations ($t^2>0$; equation\eqref{eq:tem_metal_t2gt0}) naturally predicts metallicity gradients in \hii~regions that are in good agreement with those derived from young stars. In contrast, the homogeneous case ($t^2=0$), or ``direct method'' produces a systematic offset of up to $\sim$0.3 dex, underestimating O/H abundances compared to stellar values. This discrepancy is unlikely to be explained by CNO-cycle mixing effects in stars or oxygen depletion into dust in nebulae. Contrary to the conclusions of \citet{Bresolin:16}, we find that the disagreement between nebular and stellar abundances persists even at subsolar metallicities (from 12 + log(O/H) $\sim$ 8.4).

Our nebular O/H gradient under the $t^{2}>0$ assumption also agrees with the expectations of standard Galactic chemical evolution models, indicating conventional inside–out disc formation and mild gas infall.

We also evaluated the widely used $T_{\rm e}$–metallicity relation from \citet{Shaver:83}, which assumes $t^2 = 0$ and was calibrated using pre-CCD optical data and outdated atomic coefficients. This relation yields an excessively steep metallicity gradient, significantly overestimating the O/H abundance in the inner Galaxy -- by approximately 0.2 dex at the Galactic centre -- and underestimating it by up to 0.5 dex at $R_G \sim 16$ kpc, relative to the values derived from Cepheid stars and \hii~regions analysed under the $t^2 > 0$ assumption. In light of modern CCD-based spectroscopy and updated atomic data, we strongly recommend replacing this outdated calibration with our newly derived relations, equations\eqref{eq:tem_metal_t2eq0} and \eqref{eq:tem_metal_t2gt0}, favouring the latter due to its demonstrated consistency between stellar and nebular abundances.

Lastly, we investigated the presence of azimuthal metallicity variations in the Milky Way driven by the spiral arms by analysing the residuals of the derived O/H gradients. Our results show that such variations, if they exist, are smaller than $\sim$0.1 dex, which is comparable to the typical uncertainty in the direct determination of chemical abundances using high-quality optical spectra. This suggests that the nebular gas over the extensive area sampled --covering nearly half of the Milky Way-- is relatively well mixed, indicating that the results of \citet{Esteban:22} remain valid beyond the solar neighbourhood.

\section*{Acknowledgements}

We thank the anonymous referee for a careful and constructive report that significantly improved the quality and clarity of this manuscript. JEM-D, LC, CM, MPeimbert, STP and MP  thank the support by SECIHTI CBF-2025-I-2048 project ``Resolviendo la Física Interna de las Galaxias: De las Escalas Locales a la Estructura Global con el SDSS-V Local Volume Mapper'' (PI: Méndez-Delgado). JEM-D, LC, CM, MPeimbert, STP, CE, JGR, MOG, FFRO and MP  thank the support by UNAM/DGAPA/PAPIIT/IA103326 project. JEM-D acknowledges financial support from the UNAM/DGAPA/PAPIIT/IG101025 and UNAM/DGAPA/PAPIIT/IG104325 grants. MP and JEMD acknowledge financial support from project UNAM/DGAPA/IN111423. JGR, CE and MOG acknowledge financial support from the Agencia Estatal de Investigaci\'on of the Ministerio de Ciencia e Innovaci\'on y Universidades (AEI-MCIU) under grant ``The internal structure of ionised nebulae and its effects in the determination of the chemical composition of the interstellar medium and the Universe'' with reference PID2023-151648NB-I00 (DOI:10.13039/5011000110339). JGR also acknowledges support from the AEI-MCIU and from the European Regional Development Fund (ERDF) under grant ``Planetary nebulae as the key to understanding binary stellar evolution''  with reference PID-2022136653NA-I00 (DOI:10.13039/501100011033). KK and NS gratefully acknowledge funding from the Deutsche Forschungsgemeinschaft (DFG, German Research Foundation) in the form of an Emmy Noether Research Group (grant number KR4598/2-1, PI Kreckel) and the European Research Council’s starting grant ERC StG-101077573 (“ISM-METALS").

%%%%%%%%%%%%%%%%%%%%%%%%%%%%%%%%%%%%%%%%%%%%%%%%%%
\section*{DATA AVAILABILITY}

The original data is public and available in the references cited in Section~\ref{sec:appendix}. All our calculations are present in the files of the online material. DESIRED files, although already public, can be shared upon reasonable request.



\bibliographystyle{mnras}
\bibliography{example} % if your bibtex file is called example.bib

%\begin{thebibliography}{99}
%\end{thebibliography}

%%%%%%%%%%%%%%%%%%%%%%%%%%%%%%%%%%%%%%%%%%%%%%%%%%

%%%%%%%%%%%%%%%%% APPENDICES %%%%%%%%%%%%%%%%%%%%%

\appendix

\section{Temperature, Density and Ionization Relations in Galactic \hii~Regions observed in radio}
\label{sec:Te_Ne_radio}

We have analysed the relations among electron temperature, electron density, and the ionizing-photon emission rate for the sample of Galactic \hii\ regions observed in radio. \citet{Khan:24} report all of these quantities, whereas \citet{Quireza:06} provide only temperatures and densities, with the latter lacking uncertainty estimates. \citet{Wenger:19} do not report electron densities or ionizing-photon emission rates. Our compilation of the available data is presented in Table~2 of the supplementary data.

In Fig.~\ref{fig:radio_Te_Ne}, we show a significant correlation between the electron density and the radio-derived electron temperature using the data from \citet{Khan:24}. We have chosen not to include the $n_{\rm e}$ determinations from \citet{Quireza:06} in this figure because they lack associated uncertainties and are therefore likely to be less reliable for this type of analysis. The purpose of the figure is simply to illustrate an interesting apparent correlation among the physical conditions discussed above. A more comprehensive study, in which the determination of these properties is treated in a homogeneous and statistically robust manner across all datasets, is deferred to future and more detailed work.

\begin{figure}
    \centering
    \includegraphics[width=1\columnwidth]{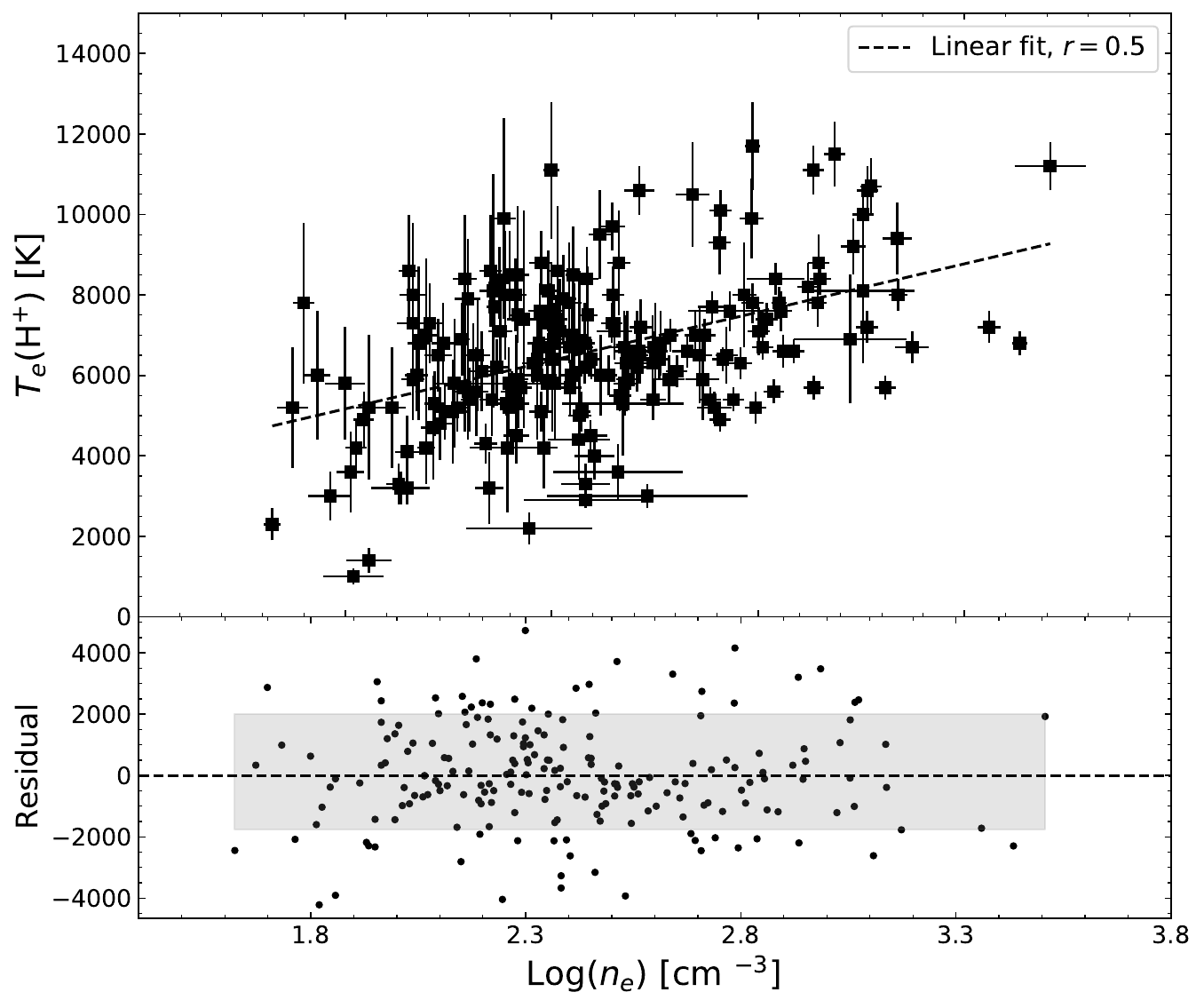}
    \caption{Relation between the electron density and electron temperature determined from radio observations for the sample of Galactic \hii~regions. The figure shows the data from \citet{Khan:24}. The quantity $r$ denotes the Pearson correlation coefficient. The shaded gray region indicates the typical fitting uncertainty $1\sigma$, corresponding to +2000K and $-$1750K}%(JEMD: RAFAEL, HAY QUE MEJORAR ESTA FIGURA. En el eje Y hay que poner Te(H+) para ser consistentes)}
    \label{fig:radio_Te_Ne}
\end{figure}

The linear fit from Fig.~\ref{fig:radio_Te_Ne} is presented in Eq.~\eqref{eq:ne_Te}:

\begin{equation}
    \label{eq:ne_Te}
    T_{\rm e}(\text{H}^{+}) = (840 \pm 795) + (2400 \pm 330)\times \log_{10} \left(n_{\rm e} \ \right)
    %Rafael, hay que incluir la ecuacion
\end{equation}

\begin{figure}
    \centering
    \includegraphics[width=1\columnwidth]{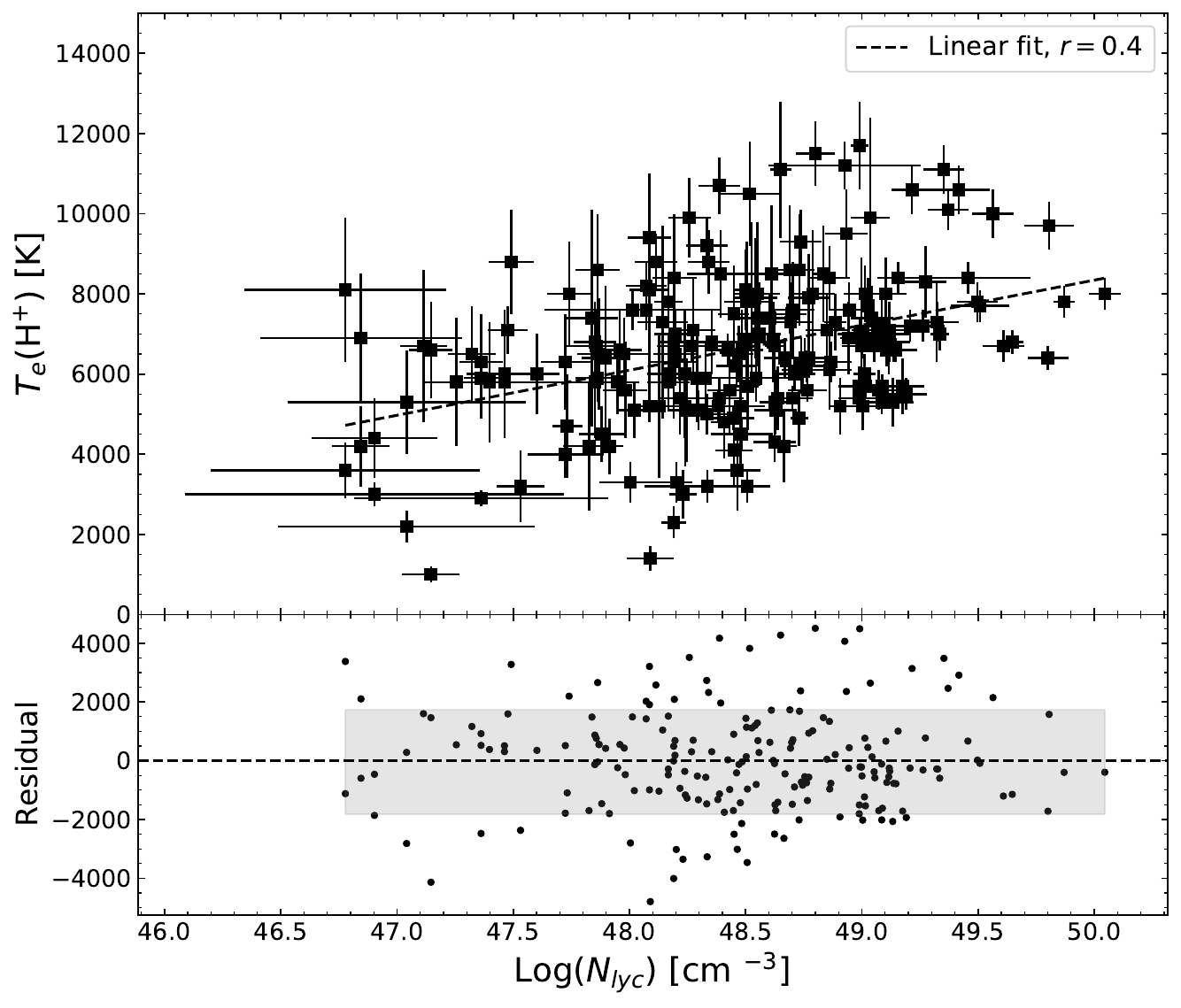}
    \caption{Relation between the ionizing-photon emission rate and electron temperature determined from radio observations for the sample of Galactic \hii~regions. The figure shows primarily the data from \citet{Khan:24}.The shaded gray region indicates the typical fitting uncertainty $1\sigma$, corresponding to +1720 K and $-$1820K.} 
    \label{fig:radio_Te_ION}
\end{figure}

Similarly, Fig.~\ref{fig:radio_Te_ION} shows the relation between the radio-derived electron temperature and the ionizing-photon emission rate, for the subset of Galactic \hii~regions from \citealt{Khan:24}. Although the scatter is larger than in Fig.~\ref{fig:radio_Te_Ne}, the correlation remains statistically significant and is presented in Eq.~\eqref{eq:Te_ion}.

\begin{equation}
    \label{eq:Te_ion}
     T_{\rm e}(\text{H}^{+}) = -(47840 \pm 9000) + (1120 \pm 190)\times \log_{10} \left(N_{\rm lyc} \ \right)
\end{equation}

We interpret Fig.~\ref{fig:radio_Te_Ne} as evidence that high electron densities can suppress part of the gas cooling through collisional de-excitation of atomic levels that would otherwise radiatively de-excite, thereby reducing the cooling efficiency. The relation between the ionizing-photon emission rate and the electron temperature shown in Fig.~\ref{fig:radio_Te_ION} is even more direct: more energetic ionizing photons produce free electrons with higher kinetic energy, which statistically leads to higher electron temperatures in the gas.

\section{Cross-Wavelength Comparison of Electron Temperatures in Galactic \hii~Regions}
\label{sec:optics_radio}
Among the Galactic \hii~regions in our sample, M~8, M~17, M~42, Sh~2-100, Sh~2-128, Sh~2-156, Sh~2-288, and Sh~2-311 have both optical observations (see Table~1 of the supplementary data) and radio measurements. These regions correspond to the radio identifiers G5.973–1.178, G015.97–00.729, G209.01–19.4, G070.280+01.583, G097.515+03.173, G110.11+0.04, G218.737+01.850, and G243.16+0.37, respectively in the radio catalogues referenced in Table~2 of the supplementary data. In the case of M42, multiple optical spectra exist, but for the purposes of this discussion we adopt the spectrum from \citet{Esteban:04} as our preferred reference.

Although the optical and radio observations partially overlap (within 1.5 arcmins), it is important to note that the apertures differ in size and are located in different regions of each nebula. Optical Galactic observations typically target small and particularly bright zones, often obtained with slit widths covering only a few arcseconds. In contrast, radio observations integrate emission over substantially larger areas of the nebulae. Given that temperature stratification varies across \hii~regions, a direct comparison between physically distinct zones of the same nebula can lead to significant differences in the inferred physical conditions.

\begin{figure}
    \centering
    \includegraphics[width=1\columnwidth]{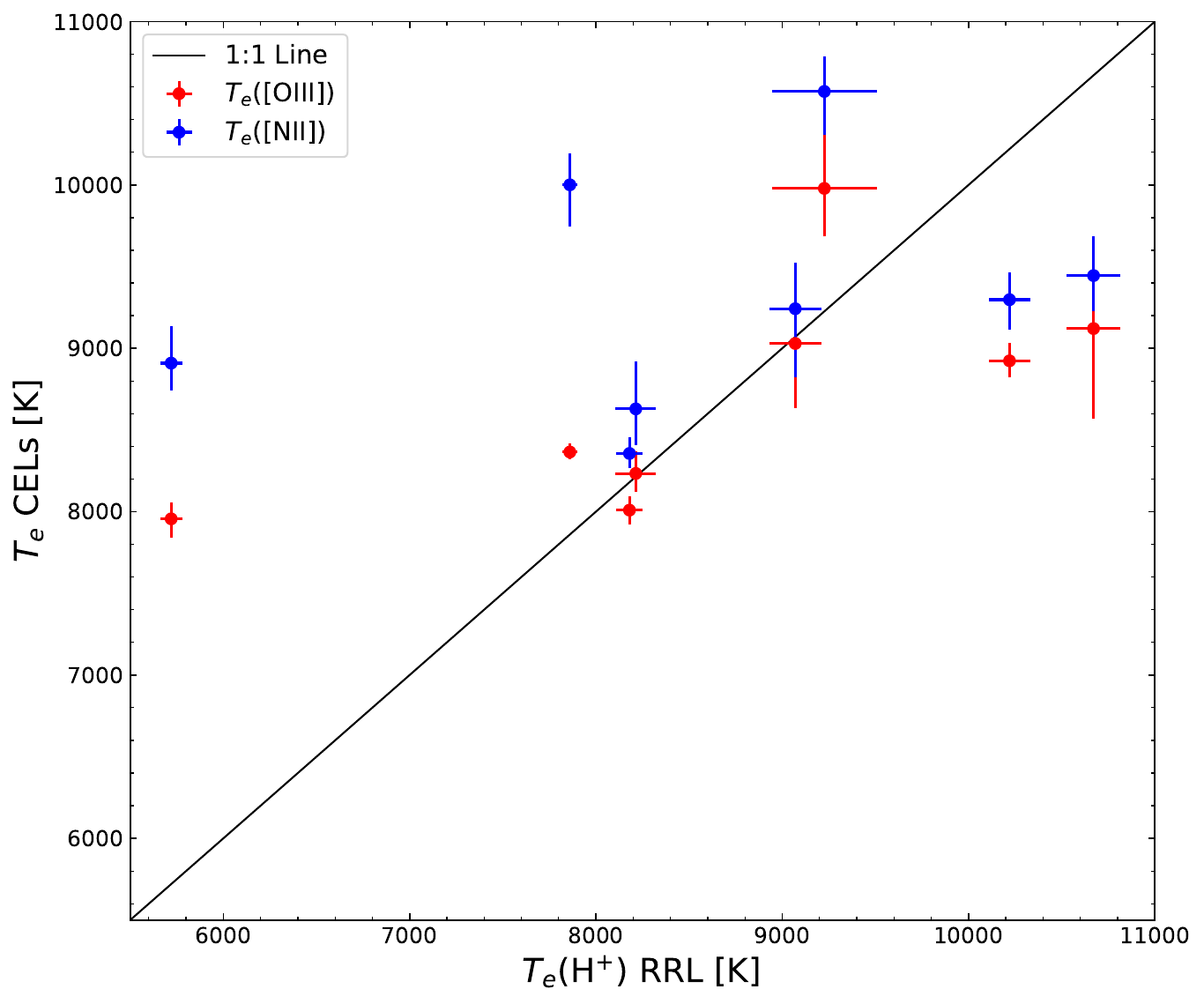}
    \caption{Direct comparison between the electron temperature derived from the nebular continuum and H I recombination detected in the radio ($T_{\rm e}$(H$^{+}$)) and the optical CEL-based temperatures $T_{\rm e}$(\oiii) and  $T_{\rm e}$(\nii) for the Galactic \hii\ regions M~8, M~17, M~42, Sh~2-100, Sh~2-128, Sh~2-156, Sh~2-288, and Sh~2-311. The matches between optical and radio observations fall within radii of up to 1.5 arcmin, and the apertures do not cover the same areas nor are they centred on the same coordinates.}
    \label{fig:opti_radio_Te}
\end{figure}

Despite these limitations, it is interesting to compare the radio-based temperatures derived from the nebular continuum and \hi~recombination with the optical temperatures obtained from CEL ratios. In principle, if internal temperature variations exist, optical diagnostics may be biased toward higher values of $T_{\rm e}$ than radio diagnostics.

In Fig.~\ref{fig:opti_radio_Te} we present the comparison between the radio-derived $T_{\rm e}(\text{H}^{+})$ and the optical determinations of  $T_{\rm e}$(\oiii) and  $T_{\rm e}$(\nii). The results are mixed: for Sh~2-288 and Sh~2-311, the radio-based $T_{\rm e}(\mathrm{H}^{+})$ is higher than the optical values, whereas in the remaining objects the agreement is good or the CEL-based temperatures appear higher. This exercise demonstrates only that the temperatures measured in the two wavelength regimes are correlated, as expected. However, the observed scatter prevents drawing firm conclusions about the presence or absence of internal temperature variations. Addressing this question requires precise control over both the aperture size and the exact position of the multiwavelength observations.

\section{References and calculations}
\label{sec:appendix}

\begin{table*}
\centering
\caption{Atomic data set used for collisionally excited lines.}
\label{table:atomic_data}
\begin{tabular}{lll}
\hline
Ion & Transition probabilities & Collision strengths \\
\hline
O$^{+}$   & \citet{FroeseFischer:04} & \citet{Kisielius:09} \\
O$^{2+}$  & \citet{Wiese:96}, \citet{Storey:00} & \citet{Aggarwal:99} \\
S$^{+}$   & \citet{Irimia:05} & \citet{Tayal:10} \\
Cl$^{2+}$ & \citet{Fritzsche:99} & \citet{Butler:89} \\
Ar$^{3+}$ & \citet{Mendoza:82b} & \citet{Ramsbottom:97} \\
Fe$^{2+}$ & \citet{Deb:09}, \citet{Mendoza:23} & \citet{Zhang:96}, \citet{Mendoza:23} \\
\hline
\end{tabular}
\end{table*}

The data used in this work were compiled from a wide range of optical and radio studies in the literature, including spectroscopic abundance determinations of ~\hii~regions and star-forming galaxies. These data were taken from \citet{Berg:13, Berg:15}, \citet{Bresolin:07a, Bresolin:09b}, \citet{Croxall:15, Croxall:16}, \citet{Esteban:04, Esteban:09, Esteban:14, Esteban:17, Esteban:20}, \citet{Fernandez:18, Fernandez:22}, \citet{GarciaRojas:04, GarciaRojas:05, GarciaRojas:06, GarciaRojas:07}, \citet{Guseva:09, Guseva:11}, \citet{Hagele:06, Hagele:08}, \citet{Izotov:04, Izotov:21b}, \citet{MendezDelgado:21a, MendezDelgado:21b, MendezDelgado:22b}, \citet{Rogers:21, Rogers:22}, as well as \citet{DelgadoInglada:16}, \citet{DominguezGuzman:22}, \citet{FernandezMartin:17}, \citet{Hirschauer:15}, \citet{Kennicutt:03}, \citet{Lin:17}, \citet{LopezSanchez:07}, \citet{Luridiana:02}, \citet{Magrini:10}, \citet{MesaDelgado:09}, \citet{Miralles-Caballero:14}, \citet{Patterson:12}, \citet{PenaGuerrero:12}, \citet{Rickards-Vaught:24}, \citet{Toribio:16}, \citet{Torres-Peimbert:89}, \citet{Tullmann:03}, \citet{Zurita:12}, together with radio observations from \citet{Quireza:06}, \citet{Wenger:19}, and \citet{Khan:24}. The complete and detailed tables summarizing the optical and radio samples are provided as supplementary material to the MNRAS.

% Don't change these lines
\bsp	% typesetting comment
\label{lastpage}
\end{document}

\label{firstpage}
\pagerange{\pageref{firstpage}--\pageref{lastpage}}
\maketitle

% ----------------------------------------------------------------------
\section*{Supplementary Material}

\onecolumn

\begin{landscape}
\renewcommand{\arraystretch}{1.3}
\scriptsize % Balanced size
\begin{longtable}{lcccccccccccccc} 
\caption{Physical conditions and ionic abundances, as well as total oxygen abundances derived assuming $t^{2}=0$ and $t^{2}>0$, expressed in units of $12 + \log(\mathrm{O/H})$ for the optical DESIRED sample. The value of $T_{\rm e}(\mathrm{H}^{+})$ was determined using the weighted averages given in Eqs.~(2) and (3) for the $t^{2}=0$ and $t^{2}>0$ cases, respectively.}
\label{table:DESIRED_data}
\\
\hline
{Galaxy} & {Region} & {$n_{\rm e}$} & {$T_{\rm e}(\text{\oiii})$} & {$T_{\rm e}(\text{\nii})$} & {O$^+$} & \multicolumn{2}{c}{O$^{2+}$} & \multicolumn{2}{c}{O} & \multicolumn{2}{c}{$T_{\rm e}(\text{H}^{+})$} & {References} \\
& & & & & &$t^2 = 0$ & $t^2 > 0$ &  $t^2 = 0$ & $t^2 > 0$ & $t^2 = 0$ & $t^2 > 0$ & \\
\hline
\endfirsthead

\multicolumn{14}{c}{{Table \thetable\ (continued)}} \\
\hline
{Galaxy} & {Region} & {$n_{\rm e}$} & {$T_{\rm e}(\text{\oiii})$} & {$T_{\rm e}(\text{\nii})$} & {O$^+$} & \multicolumn{2}{c}{O$^{2+}$} & \multicolumn{2}{c}{O} & \multicolumn{2}{c}{$T_{\rm e}(\text{H}^{+})$} & {References} \\
& & & & & &$t^2 = 0$ & $t^2 > 0$ &  $t^2 = 0$ & $t^2 > 0$ & $t^2 = 0$ & $t^2 > 0$ & \\

\hline
\endhead

%\multicolumn{13}{r}{{Continued on next page}} \\
\endfoot

\hline
\endlastfoot

HS1851p6933 & - & $470 ^{+100} _{-110}$ & $16130 ^{+330} _{-370}$ & $12200 ^{+290} _{-440}$ & $7.02 ^{+0.05} _{-0.05}$ & $7.78 ^{+0.02} _{-0.02}$ & $8.24 ^{+0.04} _{-0.03}$ & $7.85 ^{+0.02} _{-0.02}$ & $8.27 \pm 0.03$ & $15500 \pm 350$ & $15930 \pm 350$ &\citet{Izotov:21b}\\
J0021p005 & - & $140 \pm 60$ & $12490 \pm 190$ & $11060 ^{+330} _{-380}$ & $7.59 ^{+0.06} _{-0.05}$ & $7.90 ^{+0.02} _{-0.02}$ & $8.26 ^{+0.05} _{-0.04}$ & $8.07 \pm 0.02$ & $8.34 ^{+0.04} _{-0.04}$ & $12010 \pm 240$ & $12300 \pm 220$ &\citet{Hagele:06}\\
J1253-0312 & - & $420 ^{+60} _{-50}$ & $13350 ^{+100} _{-110}$ & $11950 ^{+1460} _{-1580}$ & $7.20 ^{+0.31} _{-0.15}$ & $8.01 ^{+0.01} _{-0.01}$ & $8.29 ^{+0.25} _{-0.12}$ & $8.07 \pm 0.03$ & $8.33 ^{+0.17} _{-0.17}$ & $13190 ^{+300} _{-290}$ & $13230 \pm 210$ &\citet{Thuan:05}\\
KISSR-1734 & - & $130 ^{+80} _{-70}$ & $13450 \pm 160$ & $12810 ^{+570} _{-870}$ & $7.41 ^{+0.09} _{-0.07}$ & $7.86 \pm 0.01$ & $8.04 ^{+0.06} _{-0.05}$ & $7.99 ^{+0.02} _{-0.02}$ & $8.13 ^{+0.04} _{-0.05}$ & $13280 \pm 310$ & $13310 \pm 270$ &\citet{Hirschauer:15}\\
KISSR-2117 & - & $110 ^{+70} _{-60}$ & $9600 ^{+230} _{-180}$ & $9820 ^{+630} _{-600}$ & $8.13 ^{+0.14} _{-0.10}$ & $8.03 ^{+0.03} _{-0.03}$ & $8.30 ^{+0.14} _{-0.09}$ & $8.39 ^{+0.06} _{-0.07}$ & $8.53 ^{+0.07} _{-0.08}$ & $9620 \pm 430$ & $9710 ^{+380} _{-370}$ &\citet{Hirschauer:15}\\
Mrk1236 & - & $100^{+100} _{-99}$ & $12110 ^{+110} _{-90}$ & $13120 ^{+1800} _{-1600}$ & $7.26 ^{+0.28} _{-0.14}$ & $8.08 \pm 0.01$ & $8.10 ^{+0.24} _{-0.13}$ & $8.14 \pm 0.03$ & $8.15 ^{+0.16} _{-0.15}$ & $12250 \pm 310$ & $12440 ^{+320} _{-310}$ &\citet{Izotov:04}\\
Mrk1259 & - & $790 ^{+70} _{-90}$ & $9960 ^{+440} _{-390}$ & $7960 ^{+260} _{-320}$ & $8.32 ^{+0.09} _{-0.07}$ & $7.77 ^{+0.07} _{-0.06}$ & $8.72 ^{+0.09} _{-0.07}$ & $8.43 ^{+0.06} _{-0.06}$ & $8.87 ^{+0.05} _{-0.06}$ & $8350 \pm 320$ & $9200 \pm 370$ &\citet{Guseva:11}\\
\end{longtable}

\renewcommand{\arraystretch}{1.3}
\footnotesize % Balanced size
\begin{longtable}{lccccccccccc} 
\caption{Identifications, coordinates, distances, physical conditions, and references for the Galactic \hii\ regions observed in radio. Oxygen abundances were inferred using different $T_{\rm e}(\mathrm{H}^{+})$–metallicity relations, based on the $T_{\rm e}(\mathrm{H}^{+})$ values measured from the radio data.}
\label{table:Radio_Data}

\\
\hline

Name & $\ell$ & $b$ & $d$ & $R_G$ & $\text{N}_{\text{Lyc}}$ & $n_{\rm e}$ & $T_{\rm e}(\text{H}^{+})$  & \multicolumn{3}{c}{$\text{O}$} & References \\

 & $\left[\text{°} \right]$ & $\left[\text{°} \right]$ & $\text{kpc}$ & $\text{kpc}$ & $10^{48}$s$^{-1}$ &  & $\text{K}$ & Shaver+83 & $t^2 = 0$ & $t^2 > 0$ & \\
 
\hline
\endfirsthead

\multicolumn{12}{c}{{Table \thetable\ (continued)}} \\
\hline

Name & $\ell$ & $b$ & $d$ & $R_G$ & $\text{N}_{\text{Lyc}}$ & $n_{\rm e}$ & $T_{\rm e}(\text{H}^{+})$  & \multicolumn{3}{c}{$\text{O}$} & References \\

 & $\left[\text{°} \right]$ & $\left[\text{°} \right]$ & $\text{kpc}$ & $\text{kpc}$ & $10^{48}$s$^{-1}$ &  & $\text{K}$ & Shaver+83 & $t^2 = 0$ & $t^2 > 0$ & \\
 
\hline
\endhead

%\multicolumn{13}{r}{{Continued on next page}} \\
\endfoot

\hline
\endlastfoot

G005.907-0.427 & $5.907$ & $-0.427$ & $2.18 ^{+0.58} _{-0.51}$ & $6.03 ^{+0.54} _{-0.54}$ & $1.03 \pm 0.9$ & $540 \pm {120}$ & $7600 \pm {500}$ & $8.69 \pm 0.11$ & $8.56 \pm 0.05$ & $8.72 \pm 0.08$ &\citet{Khan:24}\\
G008.662-0.342 & $8.662$ & $-0.342$ & $4.47 ^{+0.48} _{-0.39}$ & $3.84 ^{+0.42} _{-0.42}$ & $0.73 \pm 0.16$ & $310 \pm {20}$ & $5900 \pm {600}$ & $8.94 \pm 0.11$ & $8.72 \pm 0.06$ & $8.89 \pm 0.08$ &\citet{Khan:24}\\
G008.872-0.319 & $8.872$ & $-0.319$ & $4.24 ^{+0.42} _{-0.40}$ & $4.07 ^{+0.40} _{-0.39}$ & $0.95 \pm 0.24$ & $110 \pm {10}$ & $6500 \pm {900}$ & $8.85 \pm 0.15$ & $8.67 \pm 0.09$ & $8.83 \pm 0.11$ &\citet{Khan:24}\\
G010.597-0.384 & $10.597$ & $-0.384$ & $8.05 ^{+0.13} _{-0.09}$ & $1.51 ^{+0.10} _{-0.09}$ & $1.57 \pm 0.32$ & $440 \pm {30}$ & $7000 \pm {600}$ & $8.78 \pm 0.12$ & $8.62 \pm 0.06$ & $8.78 \pm 0.08$ &\citet{Khan:24}\\
G010.197-0.424 & $10.197$ & $-0.424$ & $1.69 ^{+0.45} _{-0.74}$ & $6.54 ^{+0.58} _{-0.57}$ & $0.08 \pm 0.15$ & $340 \pm {190}$ & $3000 \pm {300}$ & $9.37 \pm 0.06$ & $9.00 \pm 0.04$ & $9.17 \pm 0.06$ &\citet{Khan:24}\\
G010.142-0.422 & $10.142$ & $-0.422$ & $1.72 ^{+0.69} _{-0.49}$ & $6.51 ^{+0.60} _{-0.57}$ & $0.06 \pm 0.08$ & $290 \pm {100}$ & $3600 \pm {700}$ & $9.28 \pm 0.11$ & $8.94 \pm 0.07$ & $9.11 \pm 0.09$ &\citet{Khan:24}\\
G010.069-0.425 & $10.069$ & $-0.425$ & $2.43 ^{+0.40} _{-0.65}$ & $5.83 ^{+0.52} _{-0.50}$ & $0.29 \pm 0.22$ & $160 \pm {30}$ & $5800 \pm {900}$ & $8.96 \pm 0.15$ & $8.73 \pm 0.09$ & $8.90 \pm 0.10$ &\citet{Khan:24}\\
G010.229-0.297 & $10.229$ & $-0.297$ & $2.04 ^{+0.43} _{-0.86}$ & $6.20 ^{+0.63} _{-0.62}$ & $0.23 \pm 0.29$ & $240 \pm {80}$ & $2900 \pm {200}$ & $9.39 \pm 0.05$ & $9.01 \pm 0.03$ & $9.18 \pm 0.06$ &\citet{Khan:24}\\
\end{longtable}

\end{landscape}

\twocolumn

\bibliographystyle{mnras}
\bibliography{example} %

\label{lastpage}